# The Role of Mean Absolute Deviation Function in Obtaining Smooth Estimation for Distribution and Density Functions: Beta Regression Approach

Revised version

**Elsayed A. H. Elamir**
Department of Management and Marketing, College of Business Administration, Kingdom of Bahrain
Email: shabib@uob.edu.bh

**Abstract**

Smooth Estimation of probability density and distribution functions from its sample is an attractive and an important problem that has applications in several fields such as, business, medicine, and environment. This article introduces a simple approach but novel for estimating both functions via beta regression and generalized additive model approaches. The approach explores estimation of both functions by smoothing the first derivative of left mean absolute deviation function to obtain the final optimal smooth estimates under the condition of nondecreasing distribution function, and the density function remains nonnegative. This is achieved by using beta regression and generalized additive model with various link functions (logit, probit, cloglog, and cauchit) that applied to a polynomial function whose degree is determined by less mean absolute regression errors. Additionally, confidence limits for the distribution function are derived based on the beta distribution to give judgement about precision of obtained estimates. The method is utilized on simulated datasets featuring unimodal and multimodal and an actual dataset. The results suggest that this method exhibits strong performance relative to the kernel-based method, especially for its superior attributes in sample sizes and smoothness.

ORCID: 0000-0002-9430-072X

# 1 Introduction

Distribution and density functions have a lot of applications in several scientific domains, such as business, biology, hydrology, and environmental, among others; see, Silverman (1998), Babu et al. (2002) and Gramacki (2018). Consequently, nonparametric smoothing techniques serve as essential instruments for estimating the cumulative distribution and the density functions. These methods have evolved, significantly contributing to the expansion of nonparametric statistics as a substantial field like kernel density estimation, along with other nonparametric density estimators including Bernstein polynomial and orthogonal least squares; see, Cheng and Peng (2002), Xue and Wang (2010), Erdogan et al. (2019).

The kernel method stands out as a good technique for estimating density and probability functions. This method is fundamentally based on the principle of histogram analysis. The kernel estimate of density is

$$\hat{f}_K(x) = \left(\frac{1}{nh}\right)\sum_{i=1}^{n} K\left(\frac{x - x_i}{h}\right)$$

The distribution function is

$$\hat{F}_K(x) = \left(\frac{1}{n}\right)\sum_{i=1}^{n} K\left(\frac{x - x_i}{h}\right)$$

See; Rosenblatt (1956), Parzen (1962), Nadaraya (1964) and Silverman (1998).

A usual choice for the kernel weight $K$ is a function that satisfies $\int K(x)dx = 1$, such as Gaussain, Epanechnikov and Biweight. The performance of $\hat{F}_K(x)$ and $\hat{f}_K(x)$ is heavily dependent on the choice of bandwidth $h$. Selecting an inappropriate bandwidth can lead to either an over-smoothed estimate or under-smoothed estimates, which may not accurately reflect the underlying data distribution. Kernel estimates tend to be biased near the edges of the data distribution, especially for data with finite bounds, might require a larger dataset to produce a reliable estimate; see, see, Silverman (1986) and Gramacki (2018).

In this article, a novel approach for estimating the distribution and density functions is proposed via left mean absolute deviation (MAD) and beta regression approach. The approach finds estimation of both functions by smoothing the first derivative of left MAD function, under the conditions of nondecreasing distribution function and nonnegative density function. The derivation of these final smooth estimates is obtained by applying the beta regression of a polynomial degree $b$ on the first derivative of left MAD function. Under nonnegativity constraint of the first derivative of regression vector of expected values, the degree of polynomial is chosen as the less mean absolute regression residuals among all models. Moreover, confidence limits of the distribution function are derived using normal approximation and beta distribution. The beta regression approach has been tested on simulated data encompassing unimodal, and trimodal distributions. The results suggest that the beta regression approach performs well when compared to the kernel-based method, especially showing superior attributes in small sample sizes. This method is not intended to replace kernel but to serve as an alternative of it.

This article is organized as follows. Left MAD function representation is explained in Section 2. Beta distribution and regression are presented in Section 3. Smooth estimation of the probability and density functions is derived in Section 4. Several applications from simulated and real data are presented in Section 5. Section 6 is devoted for discussion.

## 2 Left Mean Absolute Devation

Let $X_1, \dots, X_n$ be an independent and identically a random sample from a continuous distribution function $F_X(.)$(nondecreasing and bounded $0 < F < 1$), density $f_X(.)(f \geq 0)$, mean $\mu = E(X)$, standard deviation $\sigma = \sqrt{E(x-\mu)^2}$ , indicator function $I_{i \leq k}$ (1 if $i \leq k$, 0 else), and $X_{(1)}, \dots, X_{(n)}$ are the corresponding order statistics.
The mean absolute deviation (MAD) about any value $v$ can be expressed as

$$\Delta_X(x) = E|X - x| \quad x \in R \tag{1}$$

These measures offer a direct measure of the dispersion of a random variable about its mean and median, respectively, and have many applications in different fields; see, Pham-Gia and Gorard (2005) and Elamir (2012). Munoz-Perez and Sanchez-Gomez (1990) described left MAD as

$$\Delta_X^-(x) = E[(x - X)I_{X \leq x}] \tag{2}$$

Where $I$ is the indicator function.
one big advantage of left MAD, its first and second derivative can be expressed in terms of distribution and density functions as

$$\acute{\Delta}_X^-(x) = F_X(x) \tag{3}$$

and

$$\Delta^{-''}_X(x) = f_X(x) \tag{4}$$

These equations give direct relationship between distribution and density functions and the first and second derivatives of the left MAD function, for more details; see, Munoz-Perez and Sanchez-Gomez (1990) and Elamir (2025).
It could consider nonparametric estimators of the distribution function $\acute{\Delta}_X^-(v) = F_X(v)$ using a random sample $X_1, \dots, X_n$ of size $n$ by firstly estimate left MAD function as

$$\widehat{\Delta_X^-}(v) = \frac{1}{n}\sum_{j=1}^{n}(v_i - x_j)I_{x_j \leq v_i}, \text{ for } v_i = x_{(i)}, i = 1, \dots, n$$

and its derivative as

$$\widehat{\acute{\Delta}_X^-}(x) = \hat{F}_X(x) = y_i = \frac{1}{n}\sum_{j=1}^{n} I_{x_j \leq x_i}, \text{for } i = 1, \dots, n$$

This is the empirical distribution function which has drawbacks: it has equal jumping distance $1/n$, it is not smooth as the original $F_X(x)$ and it is derivative is $1/n$.
From Elamir (2025), an estimate of first derivative of left MAD with non-equal jumping distance is given as

$$\hat{F}_a(x) = \begin{cases} \frac{1}{n}, & i = 1 \\ \frac{3i-1}{3n} - \frac{1}{3n}\frac{(x_i - x_{i-1})}{(x_{i+1} - x_{i-1})}, & i = 2, \dots, n-1 \\ \frac{n-1}{n} & i = n \end{cases}$$

This is known as forward-backward-centre estimate that uses three data points at $i-1, i,$ and $i+1$, has non equal jumping values and more efficient than empirical estimates but it is still not smoothing. If there are many ties in the data, the FC-Hermite approach can be used as

$$\hat{F}_a(x) = \begin{cases} \frac{1}{n}, & i = 1 \\ \frac{2i-1}{2n}, & i = 2, \dots, n-1 \\ \frac{n-1}{n}, & i = n \end{cases}$$

See Elamir (2025).

## 3 Beta distribution and Beta regression

The proposed model is assumed on the assumption that the response follows a beta distribution, characterized by a density that varies based on the two parameters, $p$ and $q$, which define the distribution. The expression for the beta density is

$$f(y; p, q) = \frac{\Gamma(p+q)}{\Gamma(p)\Gamma(q)} y^{p-1}(1-y)^{q-1}, \quad 0 < y < 1, p > 0, q > 0$$

$\Gamma(.)$ is the gamma function.

The mean and variance of $y$ could be written as

$$E(Y) = \frac{p}{p+q}$$

and

$$V(Y) = \frac{pq}{(p+q)^2(p+q+1)}$$

The beta distribution is highly adaptable, making it an excellent tool for modelling uncertainties across a broad spectrum of applications; see, Johnson et al. (1995).

As given in Ferrari and Cribari-Neto (2004), the beta distribution does not produce directly by the regression model. It needs re-parametrization of beta density by

$$\mu = \frac{p}{(p+q)} \text{ and } \phi = p + q$$

Where $0 < \mu < 1$ and $\phi > 0$. Therefore,

$$E(Y) = \mu, \quad var(Y) = \frac{\mu(1-\mu)}{1+\phi}$$

so that $\mu$ represents the average response, while $\phi$ serves as a measure of precision. Specifically, when $\mu$ is held constant, an increase in $\phi$ corresponds to a decrease in the variability of $Y$. The density of $Y$ can be written as

$$f(y;\mu,\phi)=\frac{\Gamma(\phi)}{\Gamma(\mu\phi)\Gamma\big((1-\mu)\phi\big)}y^{\mu\phi-1}(1-y)^{(1-\mu)\phi-1}, \qquad 0<y<1$$

Where $0<\mu<1$ and $\phi>0$.
Let $Y_i, i=1,\dots,n$ be independent random variables follows the above density with mean $\mu_i$ and precision $\phi$. By considering,

$$\mu_i=F_X(x_i)$$

The regression beta is

$$g\big(\mu_i=F_X(x_i)\big)=\sum_{j=1}^{k}x_{ij}\beta_j=X^T\beta$$

Where $\beta=(\beta_1,\dots,\beta_k)^T$ is a vector of unknown regression parameters and $x_{i1},\dots,x_{ik}$are observations on $k$ covariates $(k<n)$. Also. $g(.)$ is a strictly monotonic and twice differentiable link function that maps $(0,1)$ into $R$ . Note that the variance of $Y_i$

$$V(Y_i)=\frac{\mu_i(1-\mu_i)}{1+\phi}$$

The model inherently accommodates non-constant response variances as it is a function of $\mu_i$, which in turn depends on the values of the covariates. This allows for a flexible adaptation to varying levels of response dispersion; see, Ferrari and Cribari-Neto (2004).
There are several possible choices for the link function $g(.)$. For instance, one can use the logit function $g(\mu)=\log(\mu/(1-\mu))$, the Probit function $g(\mu)=\Phi^{-1}(\mu)$ where $\Phi^{-1}(u)$ is the cumulative distribution function of a standard normal random variable, and the complementary log-log (cloglog) link $g(\mu)=\log\{-\log(1-\mu)\}$, among others. For a comparison of these link functions, see McCullagh and Nelder (1989).
The estimated beta regression can be written as

$$\hat{g}\big(\mu_i=F_X(x_i)\big)=\sum_{j=1}^{k}x_{ij}\hat{\beta}_j=X^T\beta$$

Where $x_i^T=(x_{i1},\dots,x_{ik}), i=1,\dots,n$. The estimated beta regression is obtained by using function "betareg" from "betareg" package in R core team (2024); see, Cribari-Neto and Zeileis (2010) and "gam or bam" function in "mgcv" package in R core team (2024).

## 4 Smooth estimation

To estimate the distribution and density functions, it is assumed that

$$y_i=\hat{F}_a(x_i),\ \ i=1,\dots,n$$

Where $y$ has a beta distribution with $E(y)=\mu$ and $\text{var}(y)=\big(\mu(1-\mu)\big)/(1+\phi)$ . Moreover, it is assumed the polynomial regression from power $b$

$$y_i = \hat{F}_a(x_i) = \sum_{l=0}^{b} \beta_l x_i^l, \qquad i = 1, \dots, n$$

$\beta_l$ is the coefficients and $b$ is the power of polynomial to be chosen among estimated models that has less mean absolute residuals among nonnegative constraints for the first derivative. In this case, the function "betareg" is used from "betareg" package in R.

Alternative of "betareg" package in R is the generalized additive model (GAM). The package "mgcv" in R incorporate beta distribution inside function "bam". GAM introduced by Hastie and Tibshirani (1986) and replace $\sum_{j=1}^{k} \beta_j x_{ij}$ in regression with the function $\sum_{j=1}^{k} f_j(x_{ij})$, as

$$y_i = \hat{F}_a(x_i) = \sum_{j=1}^{k} f_j(x_{ij}), \; i = 1,2, \dots, n$$

Where $f_j$ are unknow functions, assumed to be smooth or otherwise low complexity. The function $\hat{f}_j$ is estimated in a flexible way using many smoothers such as cubic spline and B spline, and $k$ is the dimension of the basis used to represent the smooth term; see, Wood (2017).

## 4.1 Logit approach

The logit link function is an important component in logistic regression, which ensures that the predicted values fall in the range 0 and 1. It models the log-odds of the probability of an event occurring as a linear combination of independent variables, which can then be back-transformed to a probability measure; see, Wood (2017). The logit link can be written as

$$\text{logit} = \log\left(\frac{\mu}{1-\mu}\right)$$

The beta regression can be expressed as

$$g(\mu_i) = \log\left(\frac{\mu}{1-\mu}\right) = \sum_{l=0}^{b} \beta_l x_i^l = x^T \beta$$

The estimated value $\hat{\beta}$ can be obtained for different choices of $l$ via R-software R core team (2024) 'betareg' developed by Cribari-Neto and Zeileis (2010).

**Theorem 1.** The distribution function $F(x)$ can be estimated based on the beta regression (logit link) as

$$\hat{F}_{a.L}(x_i) = \hat{\mu}_i = \frac{e^{x^T\hat{\beta}}}{1+e^{x^T\hat{\beta}}} = \frac{1}{1+e^{-x^T\hat{\beta}}}$$

*Proof:* By taking the $exp$ for both sides of $g(u_i)$yields $\frac{\hat{\mu}}{1-\hat{\mu}} = e^{x^T\hat{\beta}}$ then solving with respect to $\hat{\mu}$ and divide the right-hand side by $e^{x^T\hat{\beta}}$.

**Theorem 2.** An estimate of the density function using the beta regression (logit link) is

$$\hat{f}_{a.L}(x) = \frac{d}{dx}\hat{F}_{a_1.L}(x_i) = \left(x^T\hat{\beta}\right)' \frac{e^{x^T\hat{\beta}}}{\left[1+e^{x^T\hat{\beta}}\right]^2}$$

Under condition of $\left(x^T\hat{\beta}\right)' \geq 0$, and $\left(x^T\hat{\beta}\right)' = \frac{d}{dx}\left(x^T\hat{\beta}\right)$

*Proof:* By derivative of $\frac{d}{dx}\left[e^{x^T\hat{\beta}}/\left(1+e^{x^T\hat{\beta}}\right)\right]$.

## 4.2 Probit approach

Let the $\Phi_{SN}(x)$ and $\phi_{SN}(x)$ stands for standard normal distribution and density functions, respectively. The probit link function transforms the probability of success into a linear scale; see, Wood (2017). The probit model employs the cumulative distribution function (CDF) of the standard normal distribution to estimate these probabilities. In essence, the probit link function transforms the predicted values, which can range from negative to positive infinity, to probabilities between 0 and 1, making it a powerful tool in smoothing distribution and density functions. The probit link can be written as

$$\text{Probit} = \Phi_{SN}^{-1}(\mu)$$

The beta regression can be expressed as

$$g(\mu_i) = \Phi_{SN}^{-1}(\mu) = \sum_{l=0}^{b} \beta_l x_i^l = x^T \beta$$

The estimated value $\hat{\beta}$ can be obtained for different choices of $b$ via R-software 'betareg' developed by Cribari-Neto and Zeileis (2010).

**Theorem 3.** An estimate of the distribution function based on the beta regression (probit link) is

$$\hat{F}_{a.P}(x) = \hat{\mu}_i = \widehat{\Phi}_{SN}\left(x^T \hat{\beta}\right)$$

*Proof:* Multiply both sides of $g(\mu_i)$ by $\Phi_{SN}$.

**Theorem 4.** An estimate of the density function based on the beta regression (Probit) is

$$\hat{f}_{a.P}(x) = \left(x^T \hat{\beta}\right)' f_{SN}\left(x^T \hat{\beta}\right)$$

Under condition of $\left(x^T \hat{\beta}\right)' \geq 0$ and standard normal distribution ($SN$).

*Proof:* By derivative of $\frac{d}{dx}\left[\widehat{\Phi}_{SN}\left(x^T \hat{\beta}\right)\right]$.

## 4.3 Complement loglog approach

The complementary log-log (cloglog) link function in regression is particularly useful when modelling binary outcomes where the event of interest is rare, or the data are skewed. The cloglog link is asymmetric and can model situations where the probability of the event changes more rapidly at one end of the probability scale. In essence, the cloglog link can provide a more accurate representation of the underlying process when the assumption of symmetry in the logit model is violated, leading to better model fit and more reliable estimates; see, Wood (2017). The cloglog link can be written as

$$\text{cloglog} = \log(-\log(1-\mu))$$

The beta regression can be expressed as

$$g(\mu_i) = \log(-\log(1-\mu)) = \sum_{l=0}^{b} \beta_l x_i^l = x^T \beta$$

The estimated value $\hat{\beta}$ can be obtained for different choices of $l$ via R core team (2024) 'betareg' developed by Cribari-Neto and Zeileis (2010).

**Theorem** 5. An estimate of the distribution function based on the beta regression (cloglog link) is

$$\hat{F}_{a.C}(x) = \hat{\mu}_i = 1 - e^{-e^{x^T\hat{\beta}}}$$

*Proof:* By taking $exp$ twice of $g(u_i)$.

**Theorem 6:** An estimate of the density function based on beta regression (cloglog link) is

$$\hat{f}_{a.C}(x) = \left(x^T\hat{\beta}\right)' e^{x_t^T\hat{\beta} - e^{x_t^T\hat{\beta}}}$$

Under the condition of first derivative $\left(x^T\hat{\beta}\right)' \geq 0$.

*Proof:* By derivative of $\left[1 - e^{-e^{x^T\hat{\beta}}}\right]$ with respect to $x$.

## 4.4 Cauchit approach

The Cauchit link function is particularly beneficial in generalized linear models (GLMs) when dealing with data that does not conform to the assumptions of normal distribution. The Cauchit link function, with its fewer assumptions regarding the distribution of the target, provides a more flexible approach to modelling and it can be a robust alternative to other link functions when the data shows signs of heavy tails or outliers, which might influence the model performance; see, Wood (2017).

The cauchit link can be written as

$$\text{cauchit} = \tan(\pi\mu - \pi/2)$$

The beta regression can be expressed as

$$g(\mu_i) = \tan(\pi\mu - \pi/2) = \sum_{l=0}^{b} \beta_l x_i^l$$

The estimated value $\hat{\beta}$ can be obtained for different choices of $l$ via R core team (2024) 'betareg' developed by Cribari-Neto and Zeileis (2010).

**Theorem 7.** An estimate of the distribution function based on beta regression (cauchit link) is

$$\hat{F}_{a.H}(x) = \hat{\mu}_i = \left(\frac{1}{\pi}\tan^{-1}\left(x^T\hat{\beta}\right) + \frac{1}{2}\right)$$

*Proof:* By taking the inverse of $g(u_i)$.

**Theorem 8.** An estimate of the density function based on beta regression (cloglog link) is

$$\hat{f}_{a.H}(x) = \frac{1}{\pi}\frac{\left(x^T\hat{\beta}\right)'}{\left[1 + \left(x^T\hat{\beta}\right)^2\right]}$$

Under the condition of $\left(x^T\hat{\beta}\right)' \geq 0$.

*Proof:* By derivative of $\left(\frac{1}{\pi}\tan^{-1}\left(x^T\hat{\beta}\right) + \frac{1}{2}\right)$ with respect to $x$.

## 4.5 Correction for nondecreasing

Given that the estimators $\hat{F}_{a.L}(x)$, $\hat{F}_{a.P}(x)$, $\hat{F}_{a.C}(x)$ and $\hat{F}_{a.H}(x)$ are bounded by 0 and 1, they may not exhibit a nondecreasing behaviour in the upper tail of the distribution, particularly in case of very large sample sizes. Consequently, it becomes necessary to modify $\hat{F}_X(x)$ to ensure it is monotonically nondecreasing. A viable approach for this adjustment is the bounded isotonic regression developed by Balabdaoui et al. (2009). The PAVA algorithm is employed to achieve this adjustment and is available in the R package OrdMonReg (Balabdaoui et al., 2009), specifically through the BoundedIsoMean function, as detailed by Balabdaoui et al. (2011). This function guarantees an estimate that adheres to the bounds of 0 and 1 while being monotonically nondecreasing. The modified $\hat{F}_{a_1}(x)$ is computed using the BoundedIsoMean function within the OrdMonReg package in R, resulting in a monotonically nondecreasing distribution function as

$$\hat{F}_{a_{1.h}}(x) = BoundedIsoMean\big(y = \hat{F}_a(v), w = 1/n, \mathrm{a} = 0, \mathrm{b} = 1\big), h = L, P, C \text{ and } H$$

This is nondecreasing distribution function on the range 0,1.

## 4.6 Choosing $b$ and $k$

With respect to beta regression, there are many options to choose power of polynomial $b$ from estimated models, such as, residuals, AIC and BIC criteria. To make balance between simplicity and model accuracy, the $b$ value will be chosen from estimated models that have less mean absolute error $\big((1/n)\sum_{i=1}^{n}|\hat{e}_i|\big)$ and positive first derivative $\big(x^T\hat{\beta}\big)' \geq 0$ from the range 2 to 7 as

$$b := \left(\text{less}(1/n)\sum_{i=1}^{n}|\hat{e}_i| \quad \text{subject to } \big(x^T\hat{\beta}\big)' \geq 0 \ , \qquad b = 2, \dots, 7\right)$$

With respect to GAM, the choice of $k$ will be from different estimated models that guarantee $\big(x^T\hat{\beta}\big)' \geq 0$ from the range 2 to 12 as

$$k := \left(\text{less } (1/n)\sum_{i=1}^{n}|\hat{e}_i| \quad \text{subject to } \big(x^T\hat{\beta}\big)' \geq 0 \ , \ \text{ for } k = 2, \dots, 12\right)$$

## 4.7 Properties

We show that $\hat{F}_{a_1.j}(x)$ of different links follow approximately beta distribution and derive the confidence limits for each link, by first estimating the sampling distribution of each link to allow us to use normal approximation then obtain the lower and upper confidence limits.

**Theorem 7.** Under nondecreasing and bounded of $\hat{F}_{a_1.j}(x)$ we have

(a) $E\big(\hat{F}_{a_1.j}(x)\big) = F_X(x), \quad j = L, P, C$ and $H$.

(b) $V\big(\hat{F}_{a_1.j}(x)\big) = \frac{\mu_j(1-\mu_j)}{1+\phi_j}, \quad j = L, P, C$ and $H$.

(c) $\Pr\big(\big|\hat{F}_{a_1.j}(x) - F(x)\big| > \epsilon\big) \to 0$ for $\epsilon > 0, j = L, P, C$ and $H$.

*Proof*: with respect to points (a) and (b), $\hat{F}_{a.j}(x)$ is a proportion follows approximately beta distribution, $\hat{F}_{a_1.j}(x) \sim Beta(\mu_j, \phi_j)$, $j = L, P, C$ and $H$ with mean $\mu_j = F_{a_1.j}(x)$ and variance $\frac{\mu_j(1-\mu_j)}{(1+\phi_j)}$.

For point (c), by Chebyshev's inequality $\Pr\big(\big|\hat{F}_{a_1.j}(x) - F(x)\big| > \epsilon\big) \leq \frac{V\left(\hat{F}_{a_1.j}(x)\right)}{\epsilon^2} = \frac{\mu_j(1-\mu_j)}{(1+\phi_j)\epsilon^2} \to 0$.
Note that $\phi$ is the precision parameter and inversely related to the variance, i.e, as $var \to 0$, the $\phi \to \infty$ .

Consequently, the approximate confidence limits using normal approximation for $\hat{F}$ can be written as

$$\text{Limits} = \hat{F}_{a_1.j}(x) \mp Z_{SN} \sqrt{\frac{\hat{\mu}_j\left(1-\hat{\mu}_j\right)}{\left(1+\hat{\phi}_j\right)}}$$

Since $F$ follows a $Beta\left(\mu_j, \phi_j\right)$, the confidence interval can be computed using the quantile function of beta distribution as

$$\text{Limits} = \left(Q_{\beta\left(\frac{\alpha}{2}\right)}, Q_{\beta\left(1-\frac{\alpha}{2}\right)}\right)$$

$Q$ is quantile function of beta $(\mu\phi, (1-\mu)\phi)$.

# 5 Applications

In this section we study the proposed methods using simulated and real data with comparison to kernel estimators $\hat{F}_K(x)$ and $\hat{f}_k(x)$.Two smooth bandwidths are selected using a rule of thumb bandwidth selector based on Silverman's rule (nrd0) and pilot estimation of derivatives to select the bandwidth developed by Sheather and Jones (1991) (SJ). The data is simulated using Package "distr" in R core team (2024); see, Ruckdeschel and Kohl (2014).

## 5.1 Application 1

This application is designed to study and understand the behaviour of proposed methods (betaR and GAM models) on a simulated 200 observations from standard normal distribution. The histogram for these data is given in Figure 1.

[Figure 1 about here]

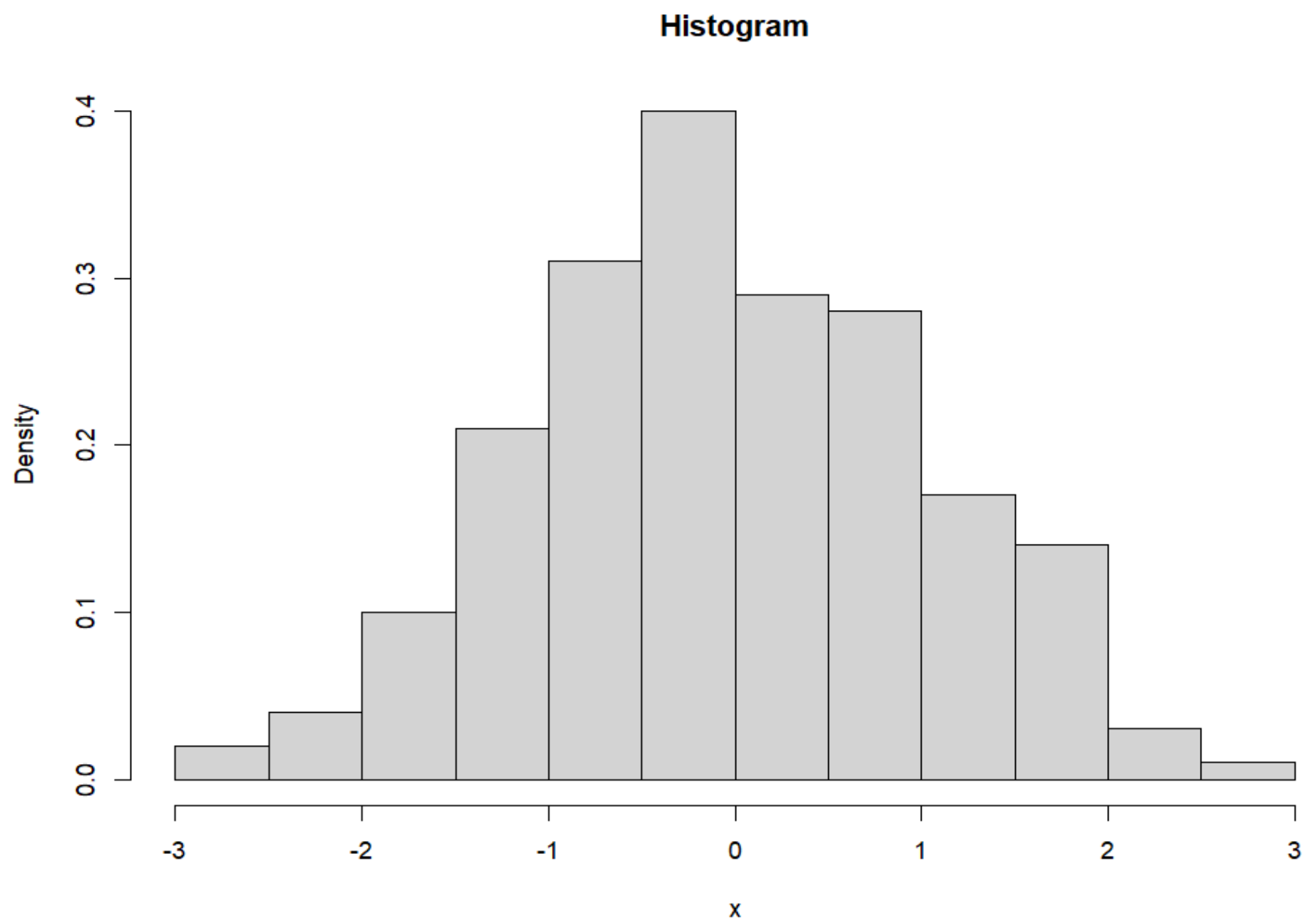

Figure1. histogram of simulated data from standard normal distribution and $n = 200$

Figure 2 shows the exact and estimated density functions using betaR with logit, probit, cloglog and cauchit links. It can be noted that the power of polynomial is 3 for logit and probit links and 4 for cloglog and cauchit links. In terms of mean absolute error, the best estimates are cauchit followed by logit, probit then cloglog. But in terms of smoothness the the estimates are probit and logit.

[Figure 2 about her]

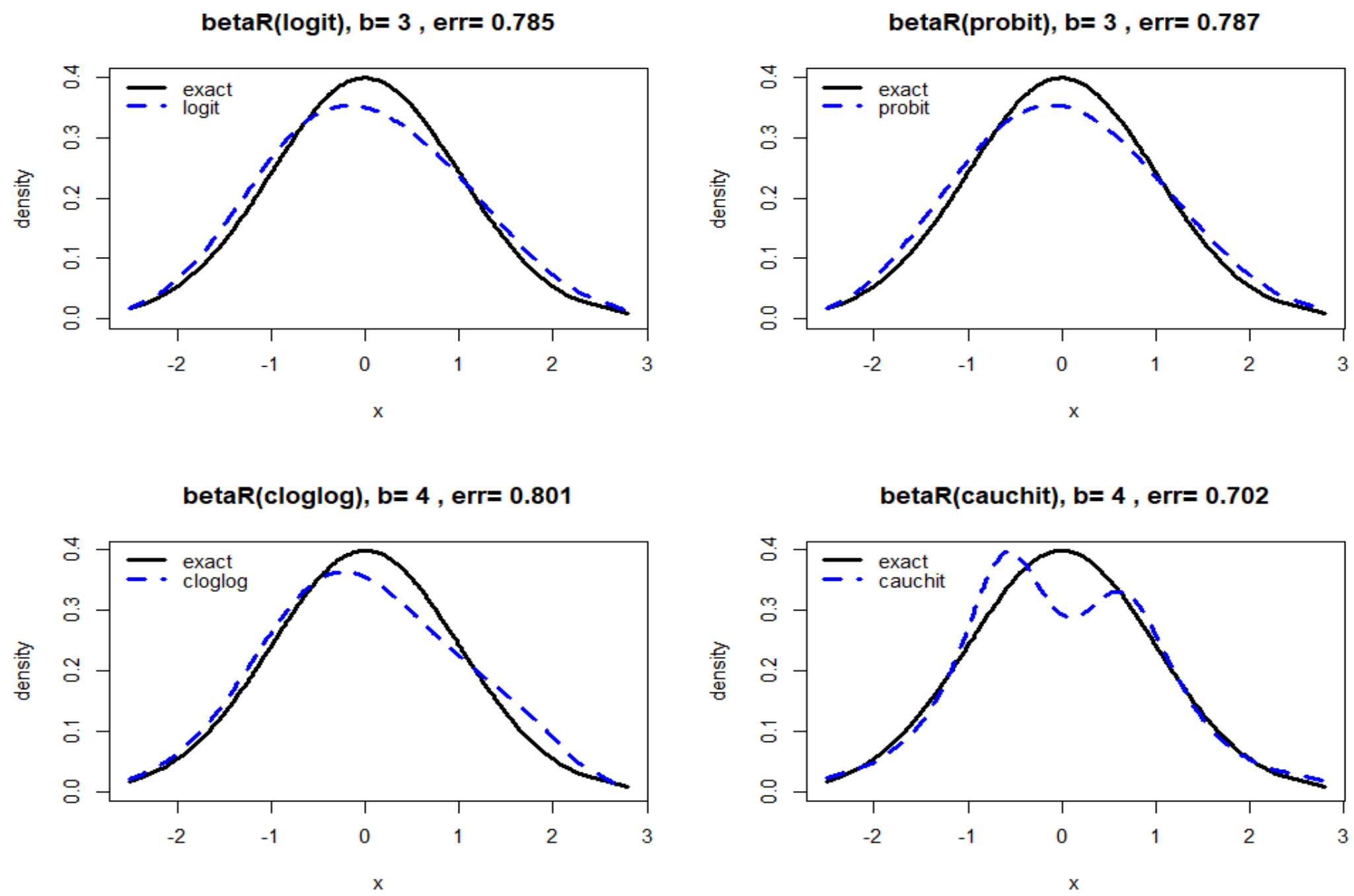

Figure 2. exact and estimated density function for simulated data from standard normal distribution using betaR and $n = 200$

Figure 3 shows the exact and estimated density functions using GAM approach with logit, probit, cloglog and cauchit links. It can be seen that the values of $k$ are 3 except cauchit 7. In terms of smoothness the probit and logit are better than others. As a result, when $k$ is smaller, the density function tends to be much smoother compared to larger value of $k$.

[Figure 3 about her]

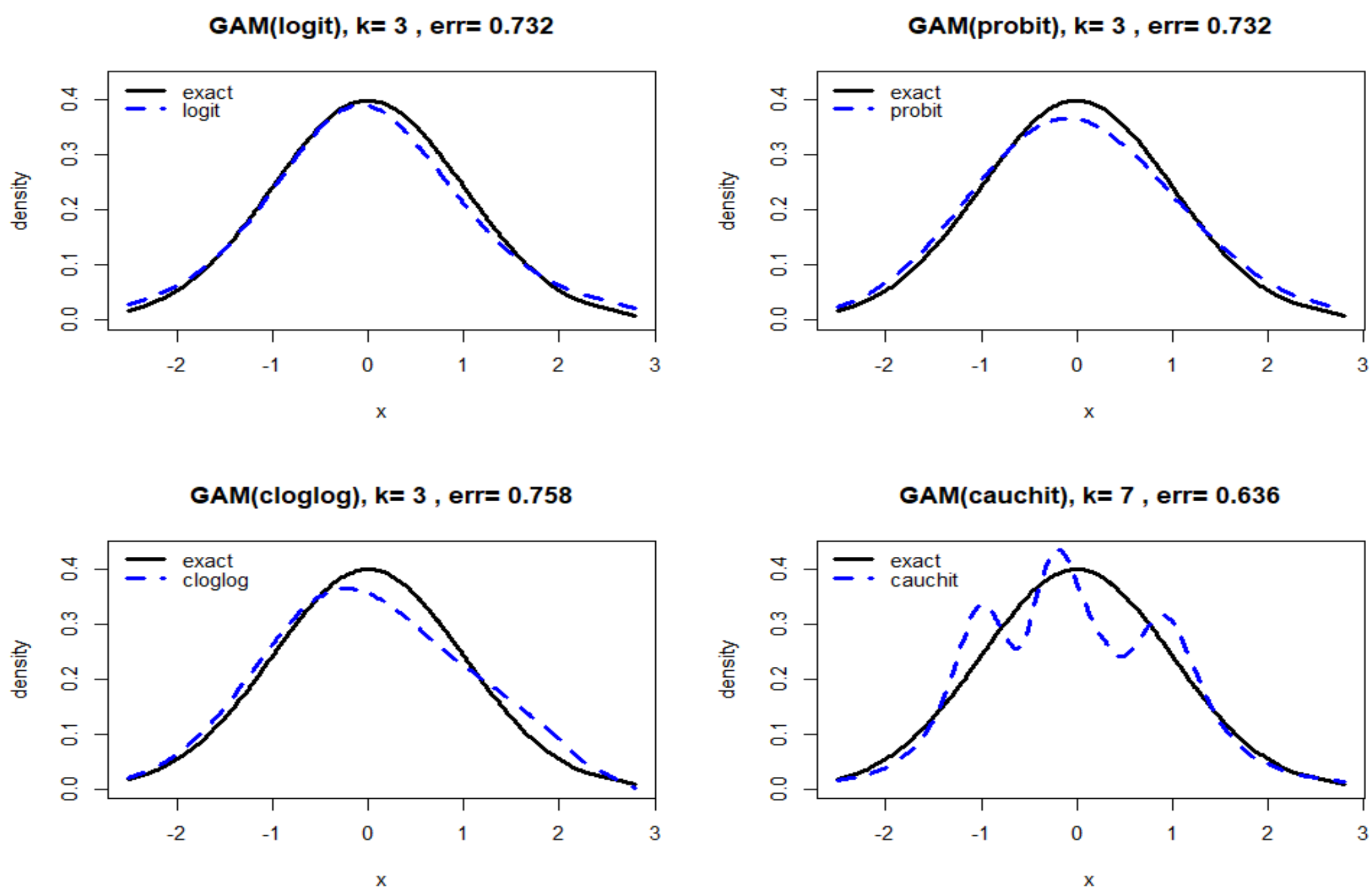

Figure 3. exact and estimated density function for simulated data from standard normal distribution using GAM and $n = 200$.

Figure 4 shows the distribution function with 99% confidence level using normal approximation and varying values of estimated precision parameter $\phi$. All the links have tight limits with high precision especially cloglog link (1532.29) indicating reliability in the estimation.

[Figure 4 about her]

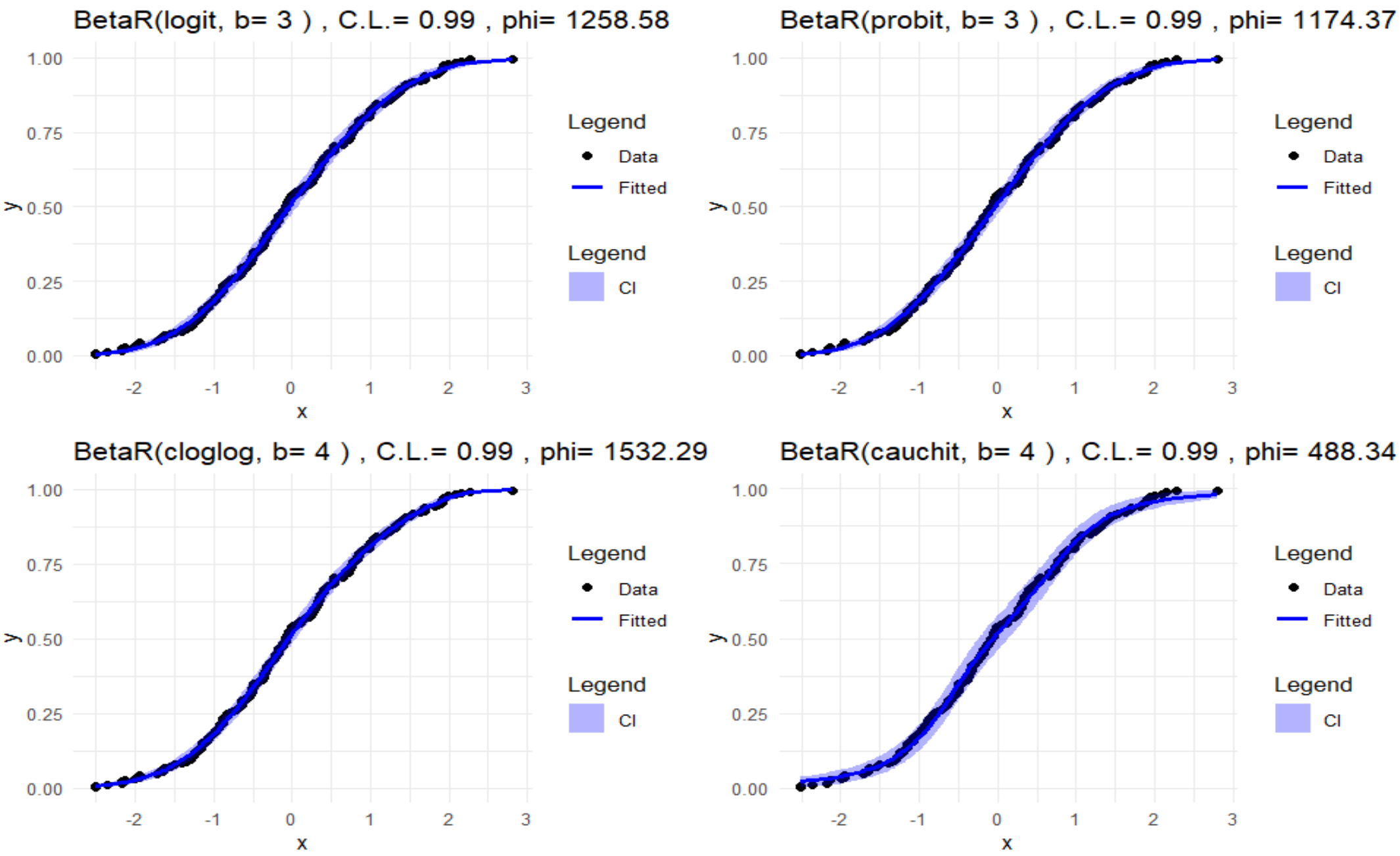

Figure 4. data, estimated distribution function with 99% confidence interval for simulated data from standard normal distribution using betaR and $n = 200$.

Figure 5 shows the distribution function with 99% confidence level using normal approximation and varying values of estimated precision parameter $\phi$. All links have tight limits with high precision especially probit link that indicating reliability in the estimation.

[Figure 5 about her]

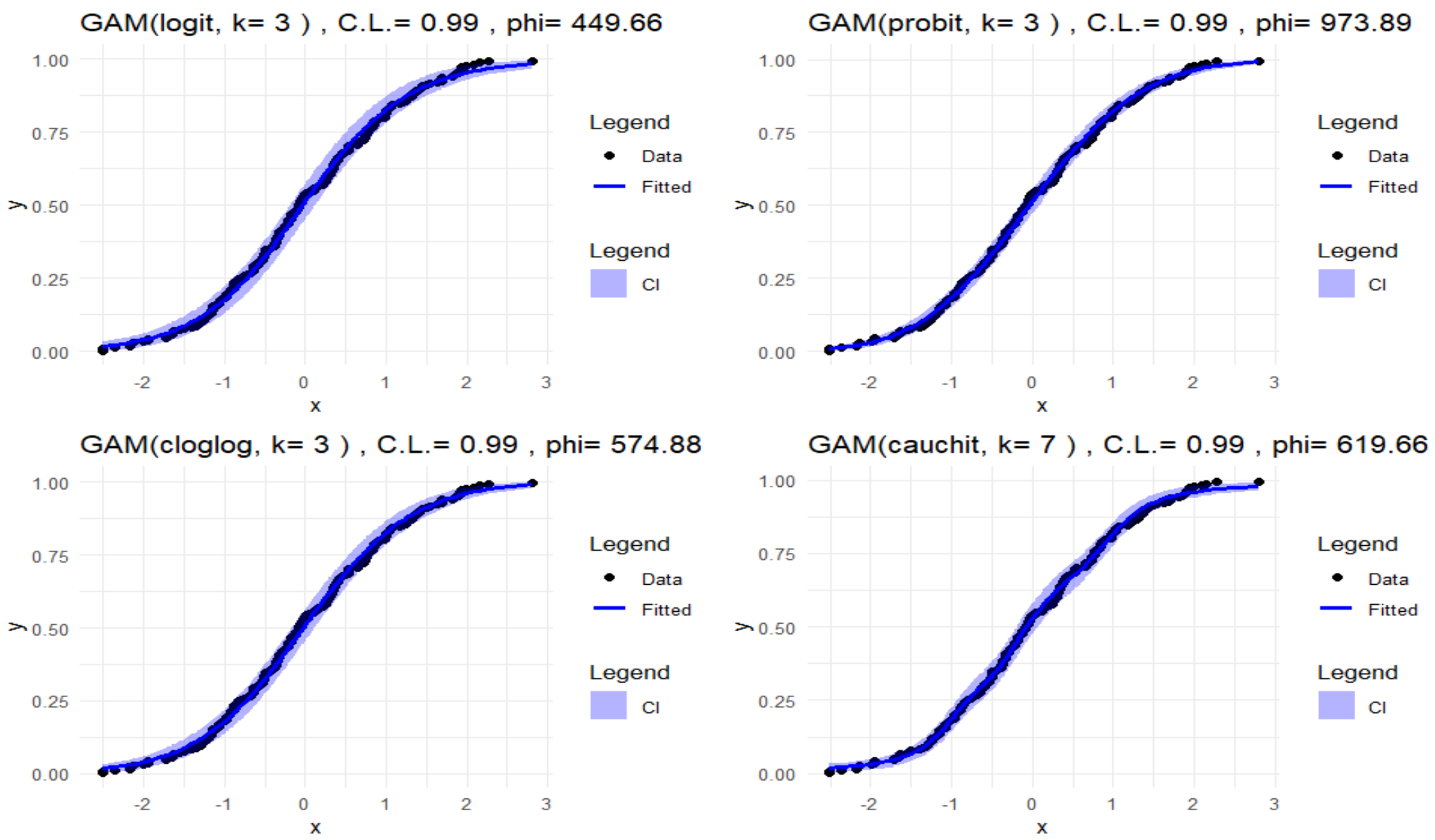

Figure 5. data and estimated distribution function with 99% confidence interval for simulated data from standard normal distribution using GAM and $n = 200$.

Figure 6 shows comparison between kernel density estimates using nrd0, SJ and GAM (logit). The graph indicates that the GAM (logit) demonstrates better fitting accuracy that making it a more effective choice for the problem at hand.

[Figure 6 about here]

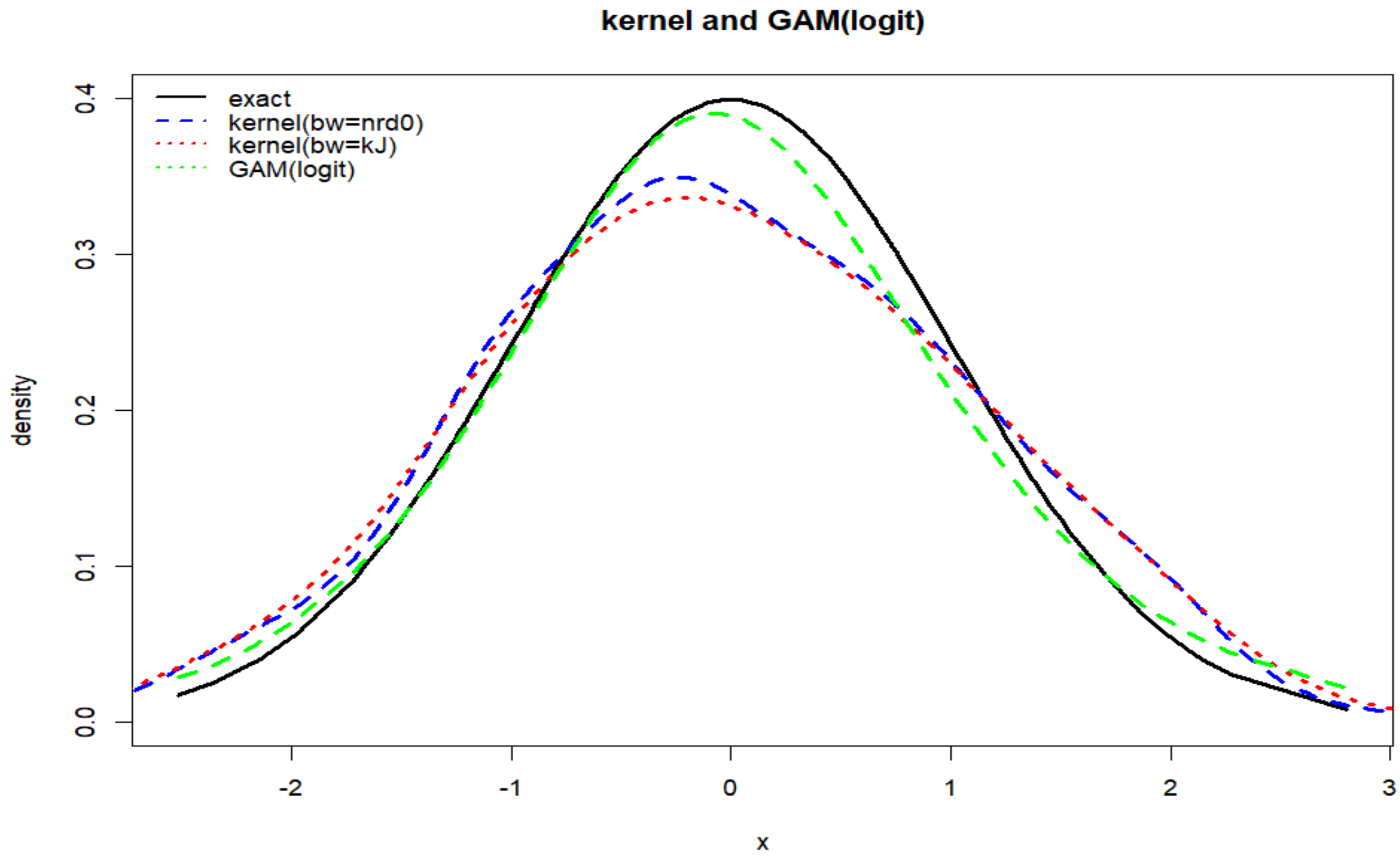

Figure 6. exact, kernel and GAM (logit) estimated density function for simulated data from standard normal distribution and $n = 200$.

## 5.2 Application 2

This application is designed to study and understand the behaviour of proposed methods (betaR and GAM approaches) on a simulated 200 observations from symmetric multimodal normal distribution. The data of size 200 is simulated from mixture normal distribution with equal probabilities $1/3N(-1, 0.20) + 1/3N(0,0.20) + 1/3N(1,0.20)$. The histogram for these data is given in Figure 7.

[Figure 7 about here]

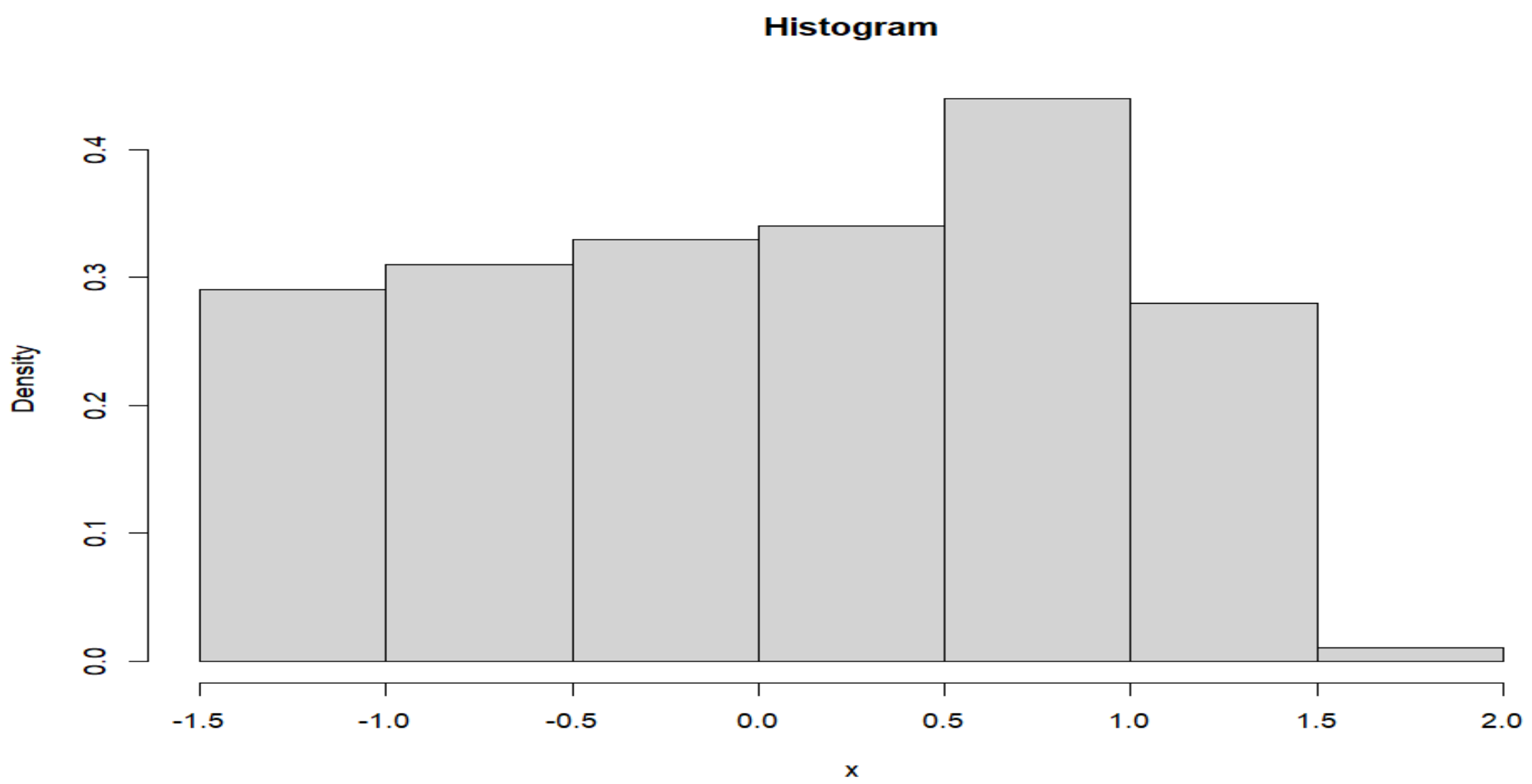

Figure 7. histogram of simulated data from mixture normal $1/3N(-1,0.2) + 1/3N(0,0.2) + 1/3N(2,0.2)$ and $n = 200$

Figure 8 shows the exact and estimated density functions using betaR with logit, probit, cloglog and cauchit links. The power of polynomial is 3 for logit and cloglog, 4 for probit and 6 for cauchit. For these data the cauchit link is much better than others in terms of mean absolute error and smoothness.

[Figure 8 about her]

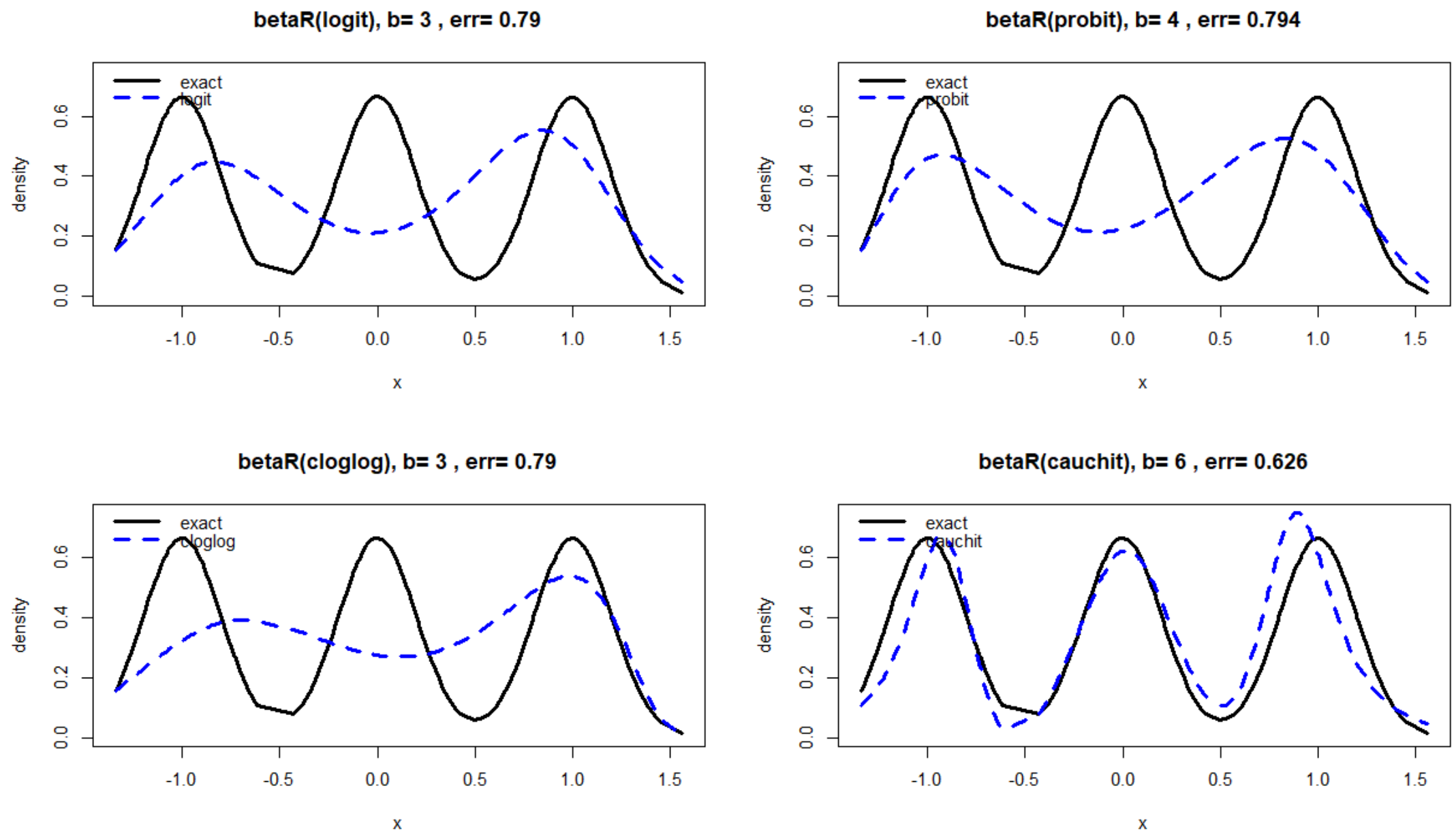

Figure 8. exact and estimated density function for simulated data from mixture normal distribution $1/3N(-1, 0.20) + 1/3N(0,0.20) + 1/3N(1,0.20)$using BetaR and $n = 200$.

Figure 9 shows the exact and estimated density functions using GAM approach with different links. Value of $k$ are 3 for logit and probit while 4 and 9 for cloglog and cauchit, respectively. In terms of smoothness and mean absolute error, cauchit link is better than others.

[Figure 9 about here]

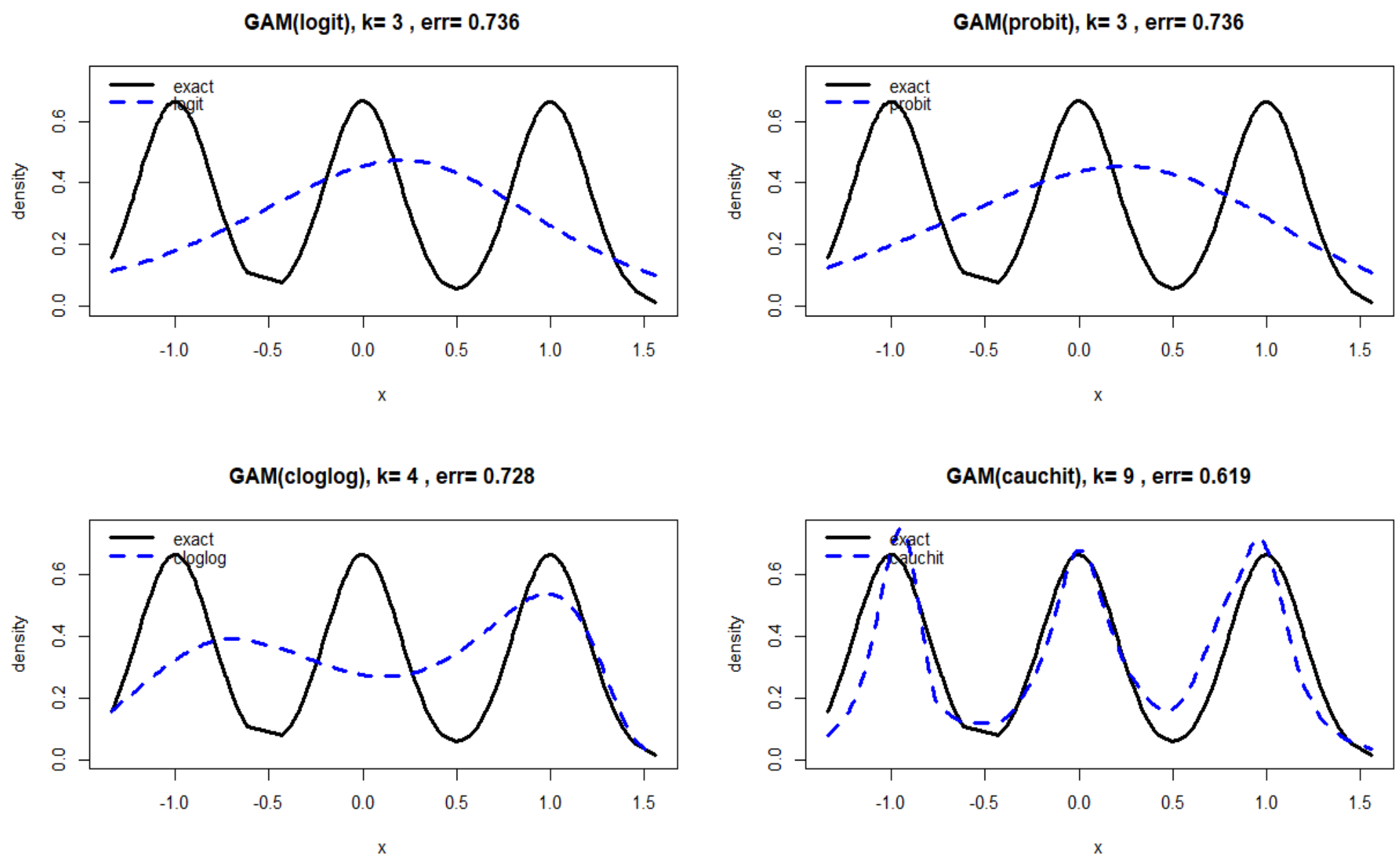

Figure 9. exact and estimated density function for simulated data from mixture normal distribution $1/3N(-1, 0.20) + 1/3N(0,0.20) + 1/3N(1,0.20)$using GAM and $n = 200$.

Figure 10 shows the distribution function with 99% confidence level of mixture normal distribution data and varying values of estimated precision parameter $\phi$. It clears that the interval is tight for bigger value of $\phi$ (precision parameter). The cauchit link has a better precision of 661.18 with tighter interval that indicating reliability in the estimation.

[Figure 10 about here]

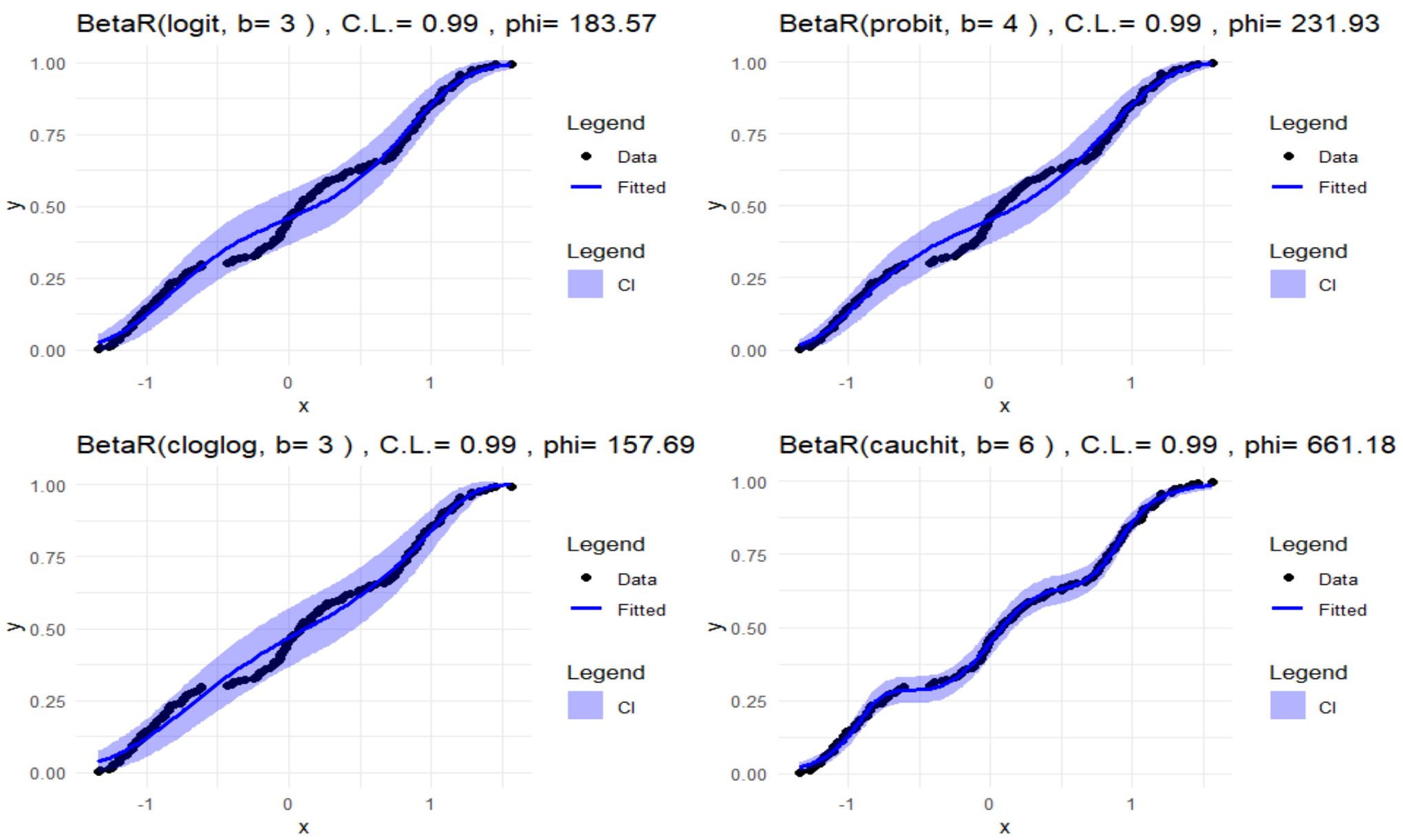

Figure 10. data, estimated distribution function with 99% confidence interval for simulated data from mixture normal distribution $1/3N(-1, 0.20) + 1/3N(0,0.20) + 1/3N(1,0.20)$ using betaR and $n = 200$.

Figure 11 shows the distribution function with 99% confidence level of mixture normal distribution data and varying values of estimated precision parameter $\phi$. Among these intervals, cauchit link illustrates better fitting due to its narrower interval when assessed against precision parameter $\phi = 573.57$.

[Figure 11 about here]

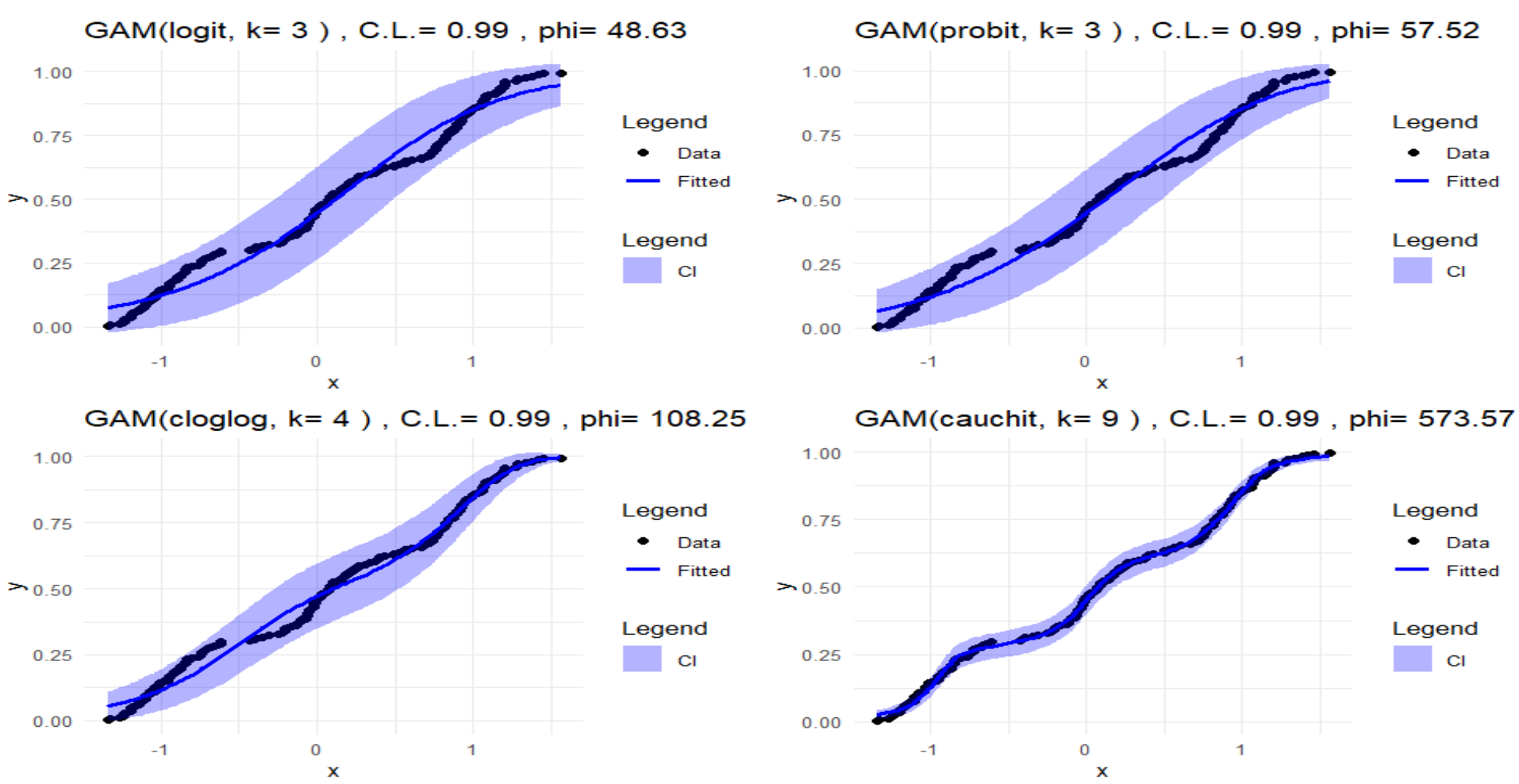

Figure 11. data, estimated distribution function with 99% confidence interval for simulated data from mixture normal distribution $1/3N(-1, 0.20) + 1/3N(0,0.20) + 1/3N(1,0.20)$ using GAM and $n = 200$.

Figure 12 shows comparison between kernel density estimates and GAM (cauchit). The graph indicates that the GAM(cauchit) demonstrates slightly better fitting accuracy making it a good alternative choice for the problem at hand.

[Figure 12 about here]

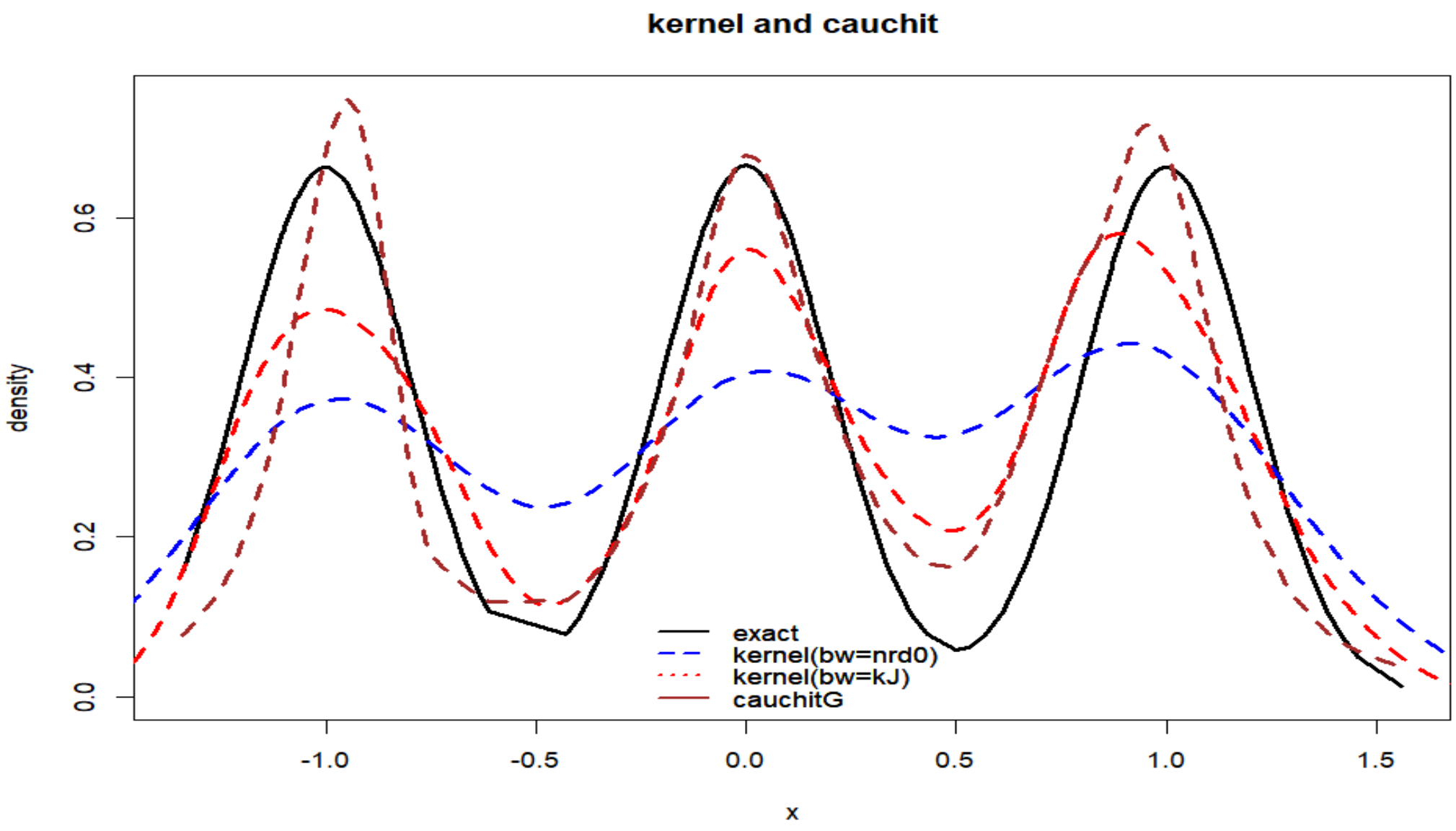

Figure 12. exact, kernel(SJ) and GAM (cauchit) estimated density function for simulated data from mixture normal distribution $1/3N(-1, 0.20) + 1/3N(0,0.20) + 1/3N(1,0.20)$ and $n = 200$.

## 5.3 Application 3

This application is designed to study and understand the behaviour of proposed methods (betaR and GAM approaches) on a simulated 200 observations from asymmetric multimodal normal distribution. The data of size 200 is simulated from mixture normal distribution with nonequal probabilities $5/10N(-1, 0.20) + 3/10N(0,0.20) + 2/10N(1,0.20)$. The histogram for these data is given in Figure 13.

[Figure 13 about here]

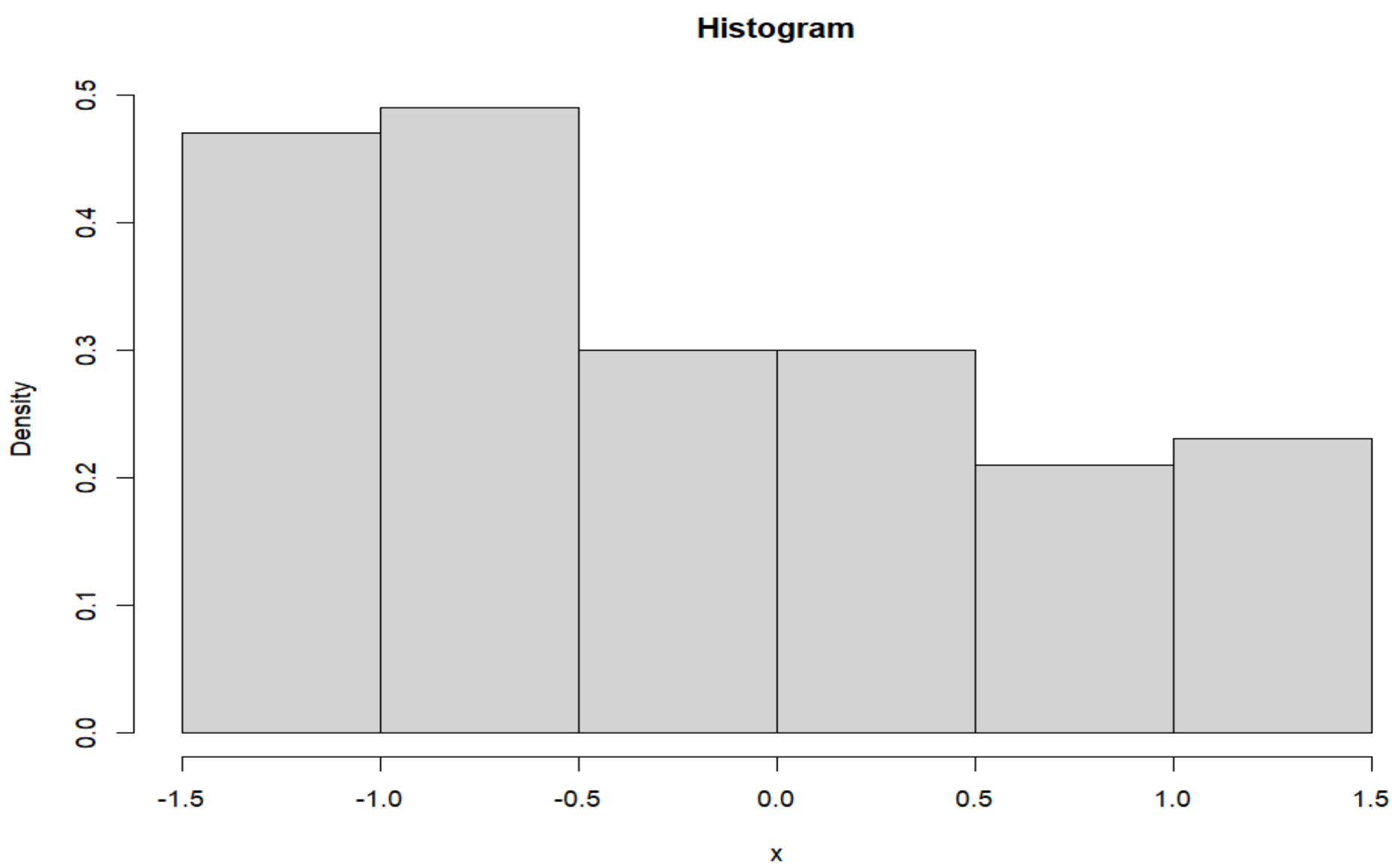

Figure 13. histogram of simulated data from mixture normal $5/10N(-1,0.2) + 3/10N(0,0.2) + 2/10N(2,0.2)$ and $n = 200$

Figure 14 shows the exact and estimated density functions using betaR with logit, probit, cloglog and cauchit links. The power of polynomial is 5 for logit, 6 for cloglog, 7 for probit and 6 for cauchit. For these data the cauchit link is much better than others in terms of mean absolute error.

[Figure 14 about her]

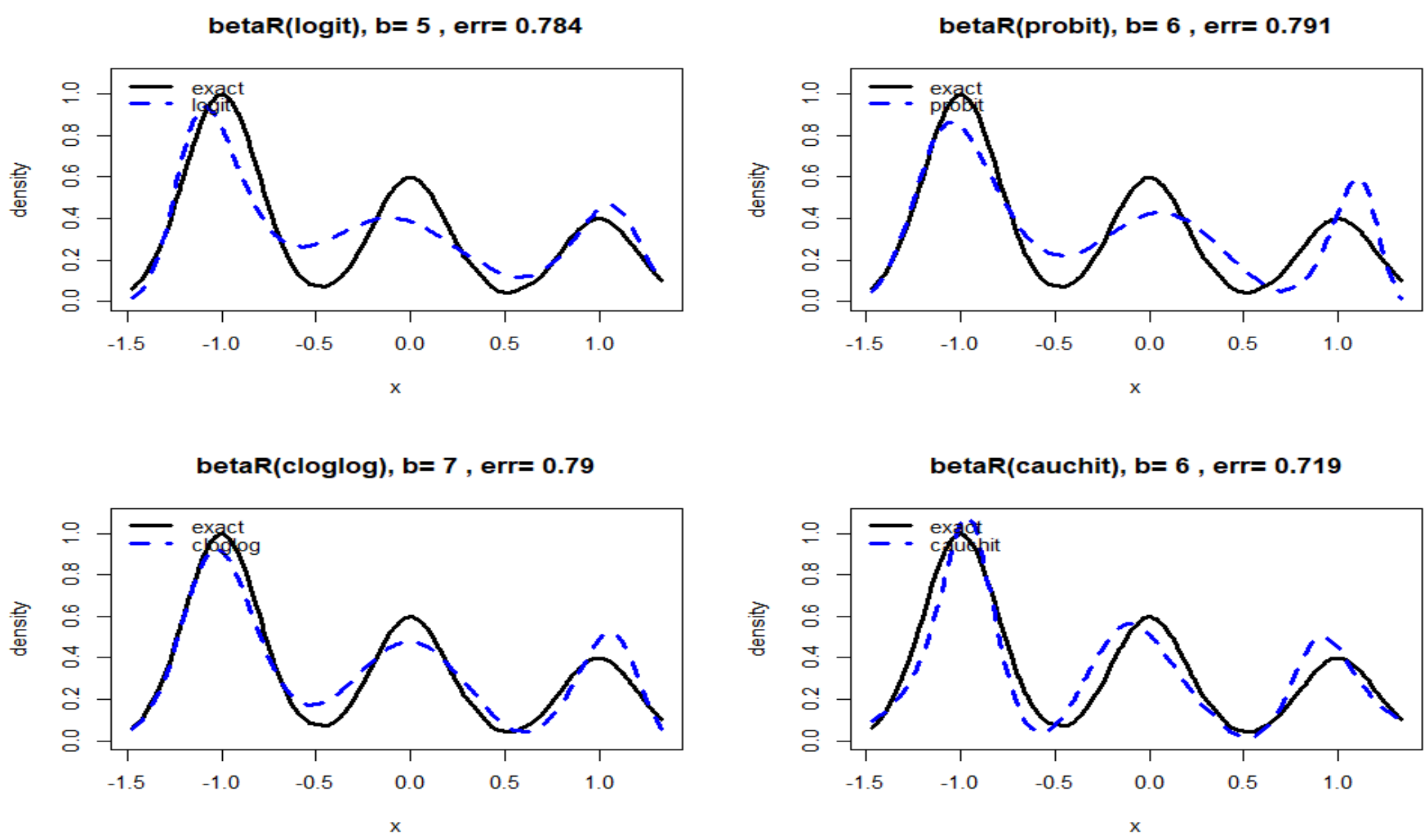

Figure 14. exact and estimated density function for simulated data from mixture normal distribution $5/10N(-1, 0.20) + 3/10N(0,0.20) + 2/10N(1,0.20)$using BetaR and $n = 200$.

Figure 15 shows the exact and estimated density functions using GAM approach with different links. Values of $k$ are 5 for logit and probit while it is 8 and 12 for cloglog and cauchit, respectively. In terms of mean absolute error, the cauchit link is better than others.

[Figure 15 about here]

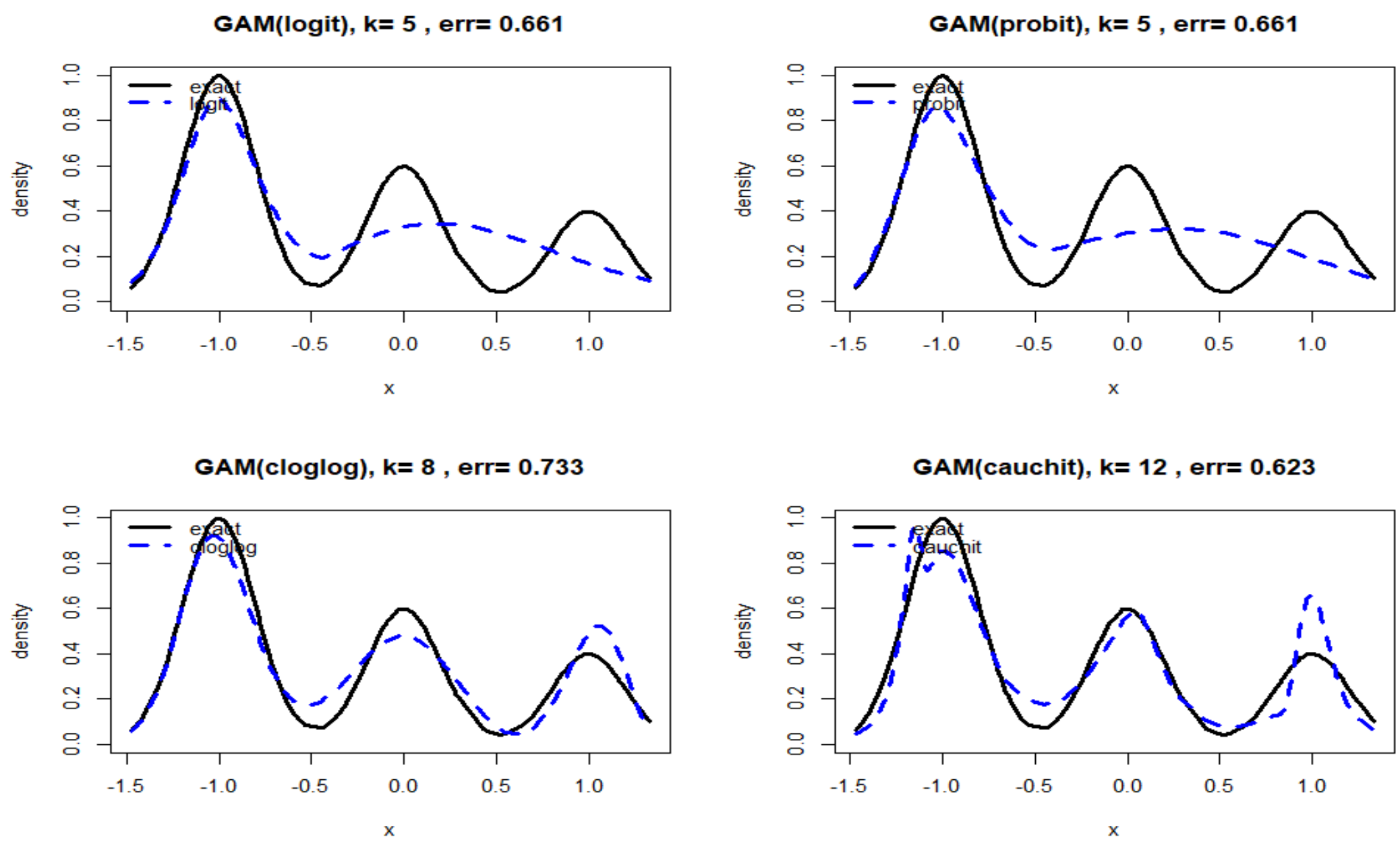

Figure 15. exact and estimated density function for simulated data from mixture normal distribution $5/10N(-1, 0.20) + 3/10N(0,0.20) + 2/10N(1,0.20)$using GAM and $n = 200$.

Figure 16 shows the distribution function with 99% confidence level of mixture normal distribution data and varying values of estimated precision parameter $\phi$. The cloglog and probit links have better precision 2095 and 1225.96 with tighter interval that indicating reliability in the estimation.

[Figure 16 about here]

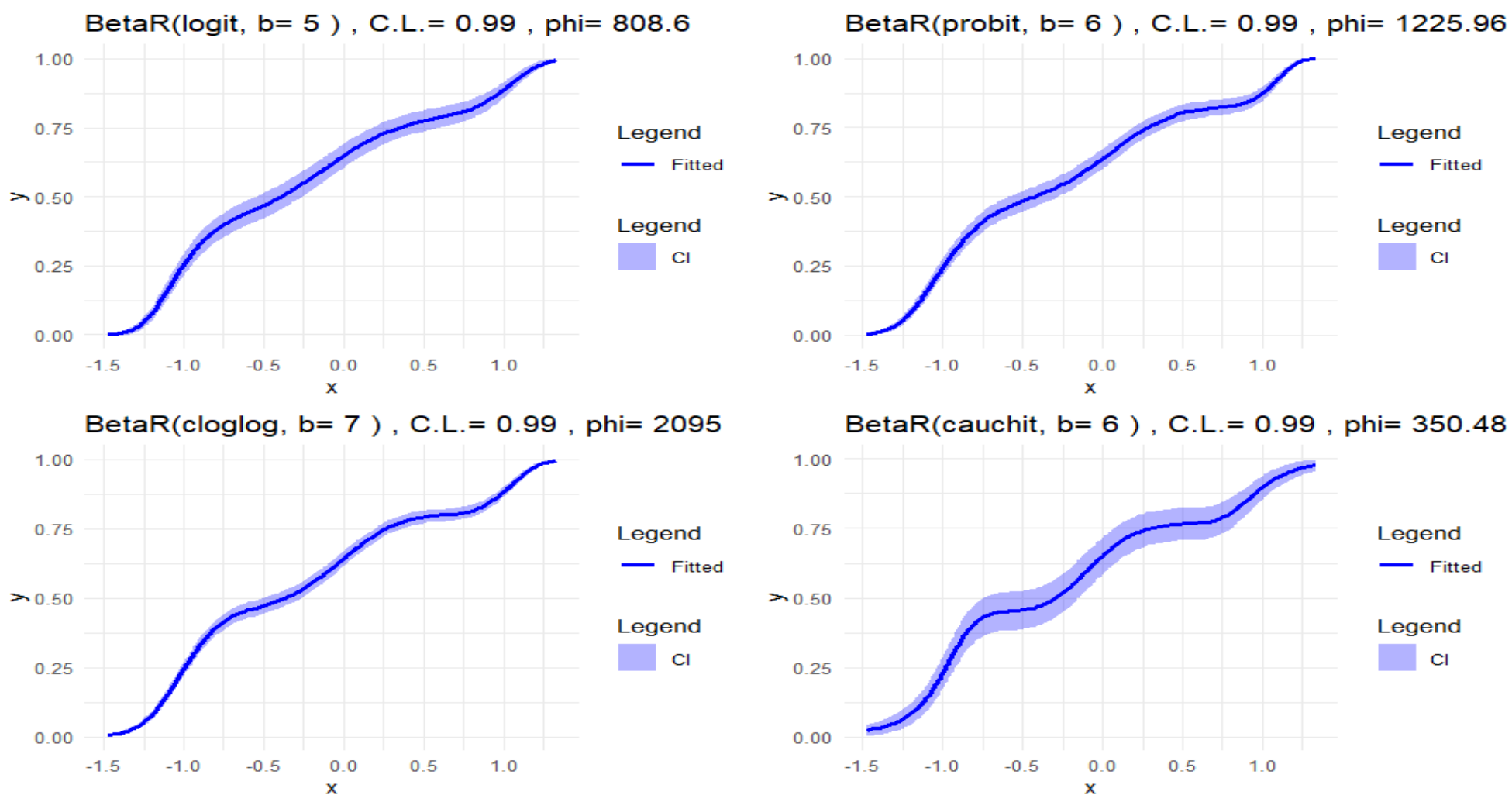

Figure 16. estimated distribution function with 99% confidence interval for simulated data from mixture normal distribution $1/3N(-1, 0.20) + 1/3N(0,0.20) + 1/3N(1,0.20)$ using betaR and $n =$ 200.

Figure 17 shows the distribution function with 99% confidence level of normal mixture distribution data and varying values of estimated precision parameter $\phi$. Among these intervals, cloklog and cauchit links illustrate better fitting due to its narrower interval when assessed against precision parameters $\phi = 2264.65$ and $1146.57$.

[Figure 17 about here]

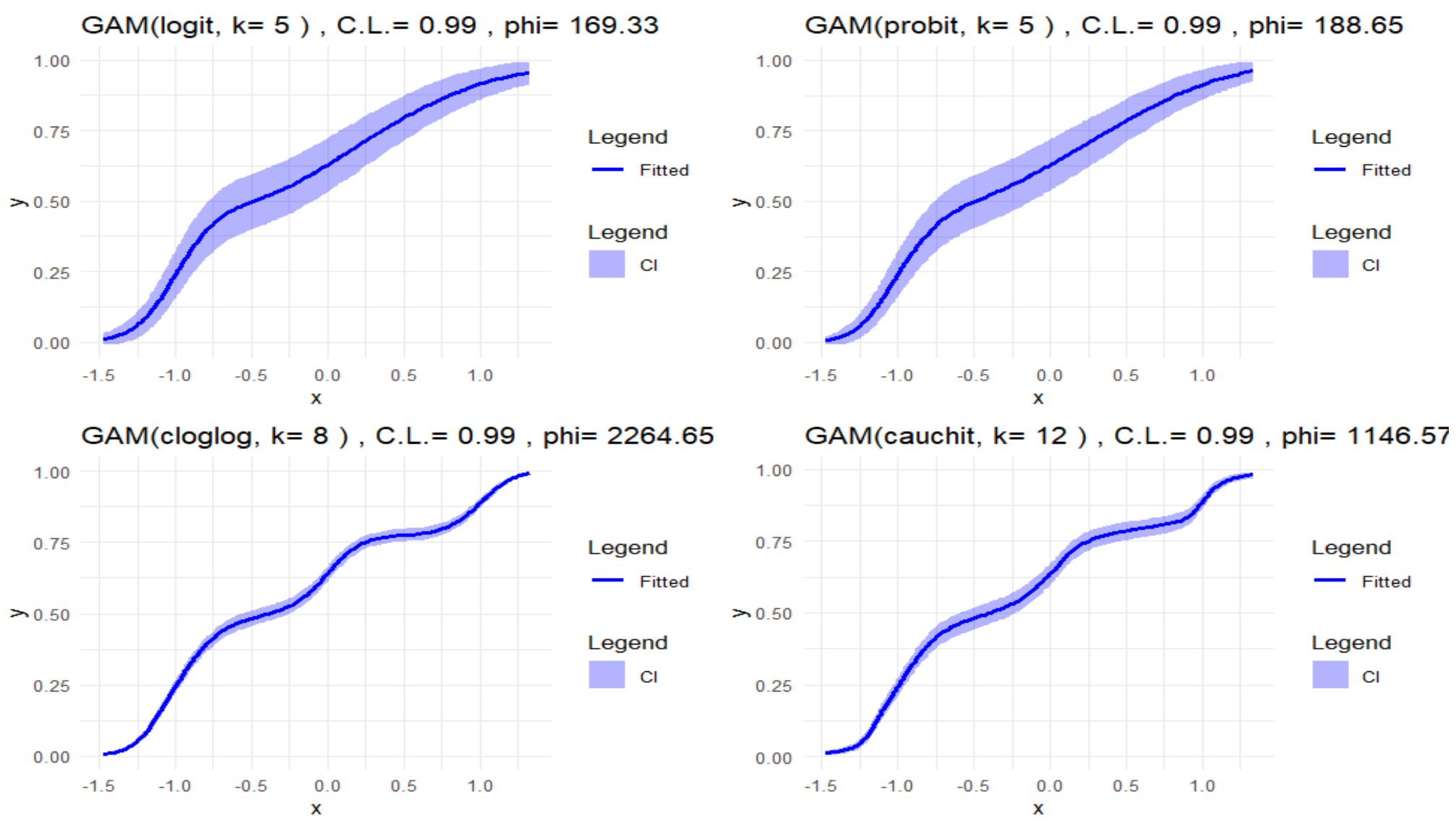

Figure 17. estimated distribution function with 99% confidence interval for simulated data from mixture normal distribution $1/3N(-1, 0.20) + 1/3N(0,0.20) + 1/3N(1,0.20)$ using GAM and $n =$ 200.

Figure 18 shows comparison between kernel density estimates using bw=nrd0, SJ and GAM (cauchit). The graph indicates that the GAM (cauchit) demonstrates slightly better fitting accuracy that making it a good alternative choice for the problem at hand.

[Figure 18 about here]

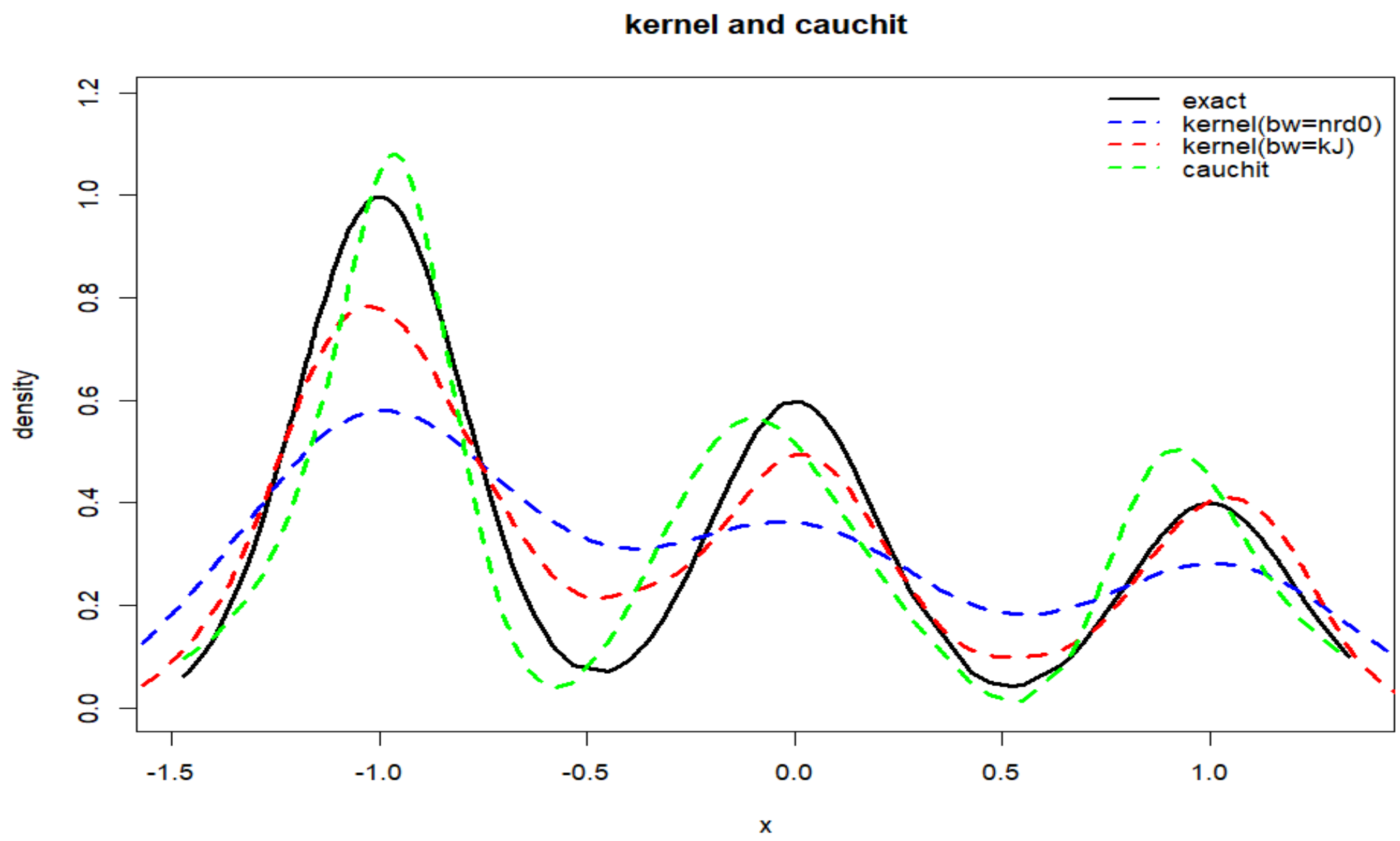

Figure 18. exact, kernel and GAM (cauchit) estimated density function for simulated data from mixture normal distribution $5/10N(-1, 0.20) + 3/10N(0,0.20) + 2/10N(1,0.20)$ and $n = 200$.

## 5.4 Real data

A dataset comprising 2276 records on baseball team batting statistics has been acquired from the following source: 'https://raw.githubusercontent.com/hillt5/DATA_621/master/HW1". The objective is to analyse the dataset distribution and determine the presence of bimodality. The histogram in Figure 19 illustrates the distribution of team batting home runs (HR). To ascertain whether the distribution is unimodal, Hartigans' dip test can be employed using the dip.test() function from the diptest package; see, Maechler (2021) and the R core team (2024). According to Hartigans' dip test for unimodality, the diptest yields a p-value less than 0.01, suggesting the distribution is not unimodal and may be bimodal or multimodal.

[Figure 19 about here]

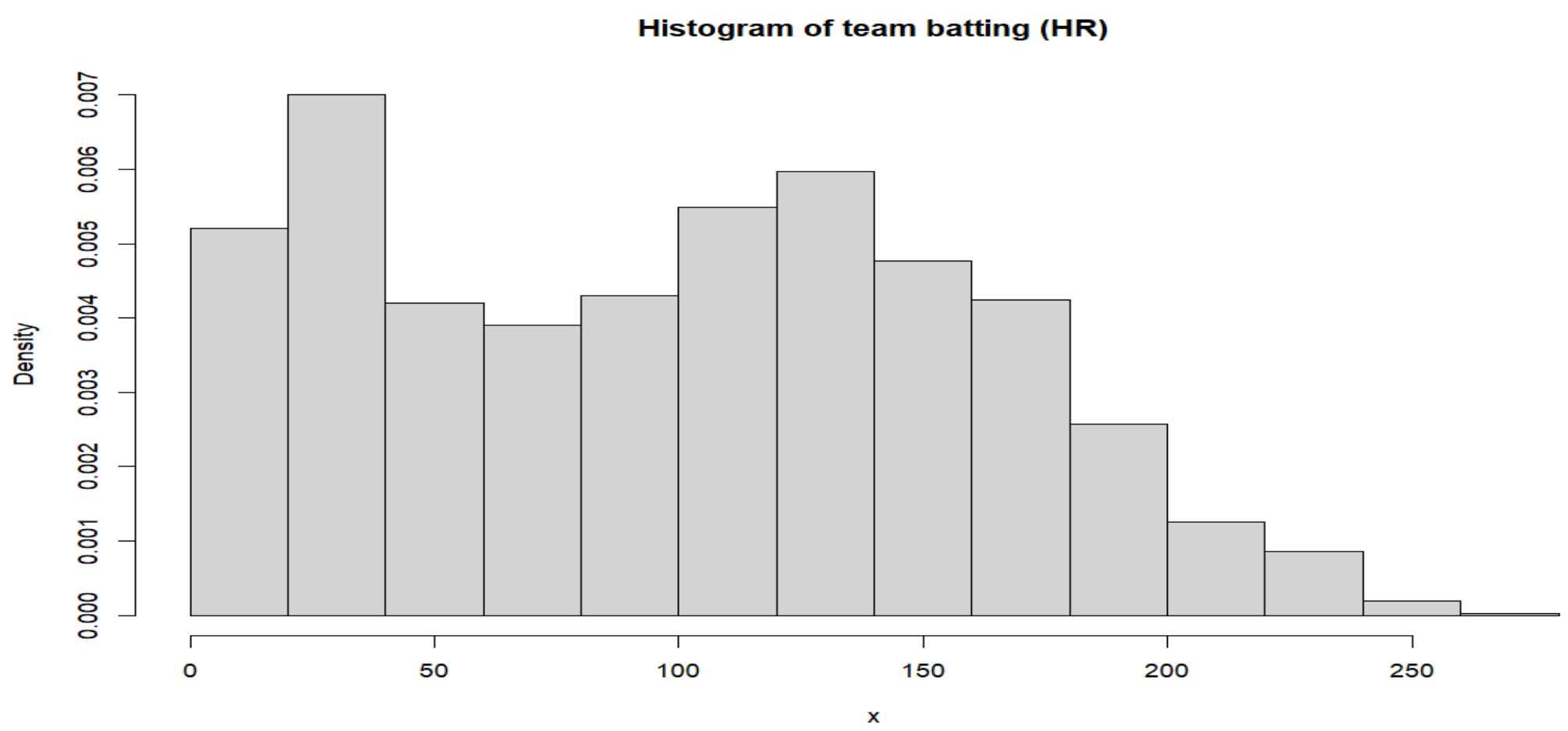

Figure 19. Histogram of team batting data

The estimating distribution functions with confidence level based on several methods are given in Figure 20. The models logit, probit and cauchit necessitate polynomials of degree 5, while the GAM (logit, probit and cauchit) has $k = 8$, 8, 12, respectively. BetaR (probit) give the best precision $\phi$ (6227.52).

[Figure 20 about here]

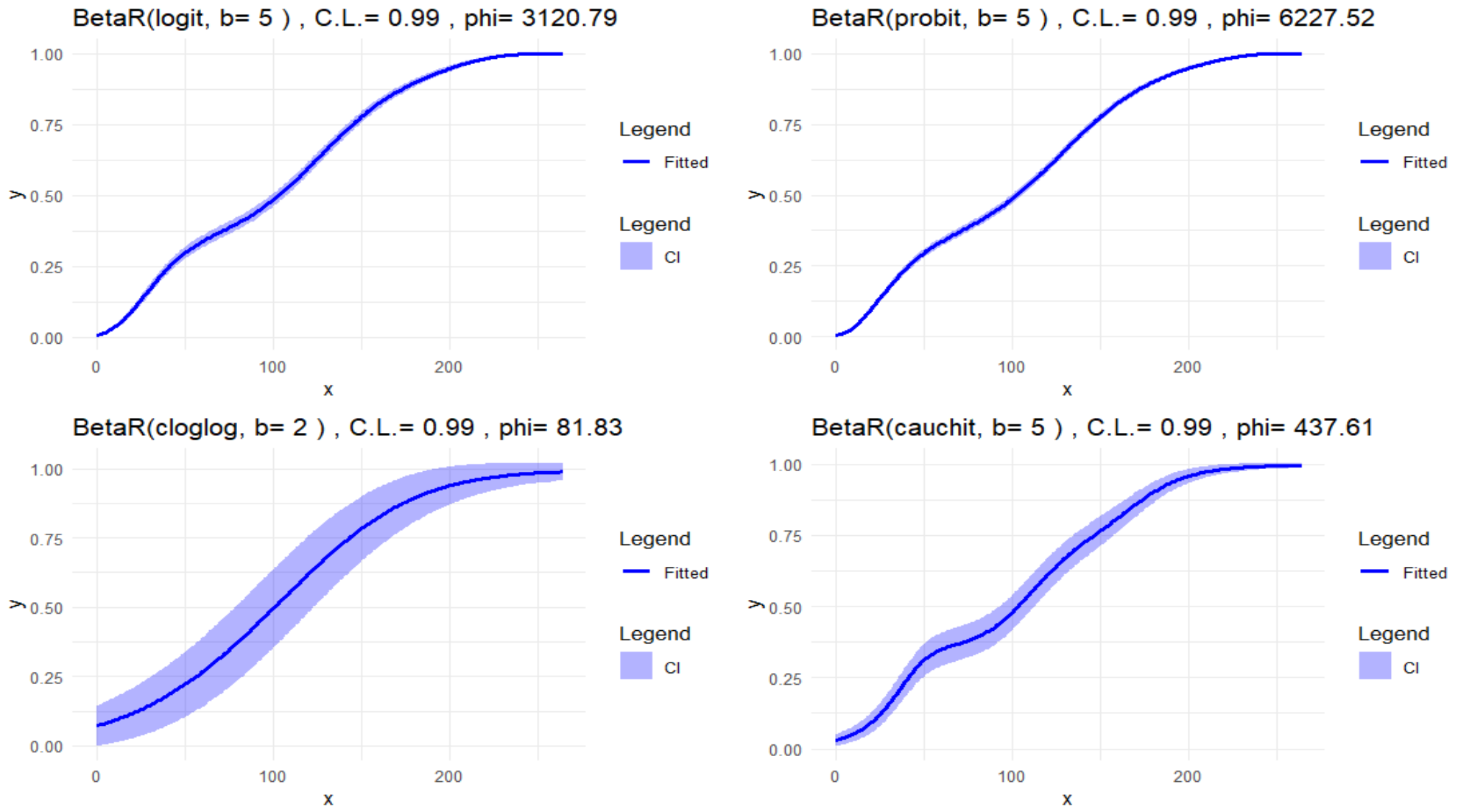

Figure 20. estimated distribution function with 99% confidence interval for 2276 records on baseball team batting using BetaR approaches.

[Figure 21 about here]

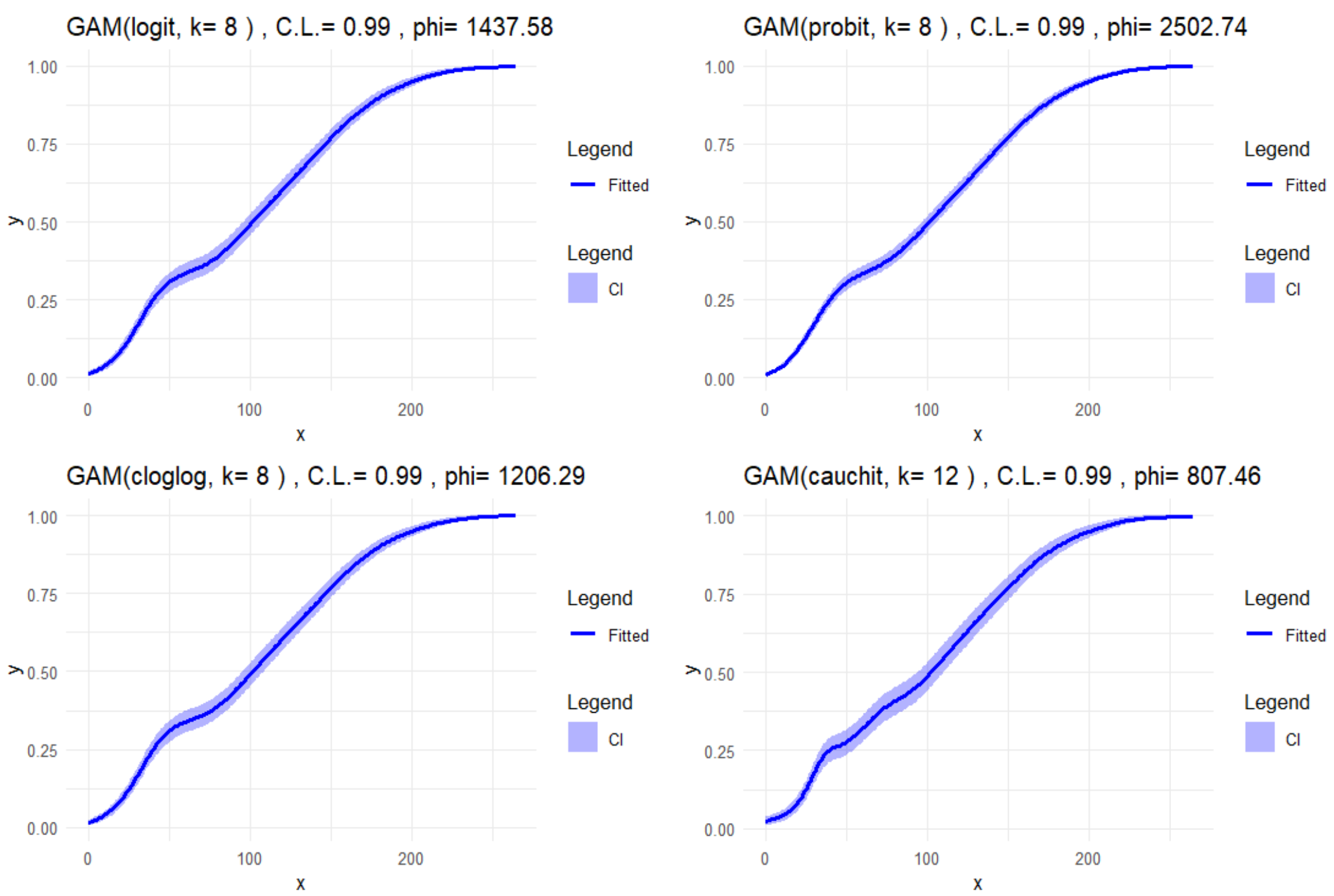

Figure 21. estimated distribution function with 99% confidence interval for 2276 records on baseball team batting using GAM approaches.

Figure 22 shows comparison between kernel density estimates (SJ) and BetaR (probit). The graph indicates that both of them demonstrate a good fitting for the data, but BetaR shows better smoothness.

[Figure 22 about here]

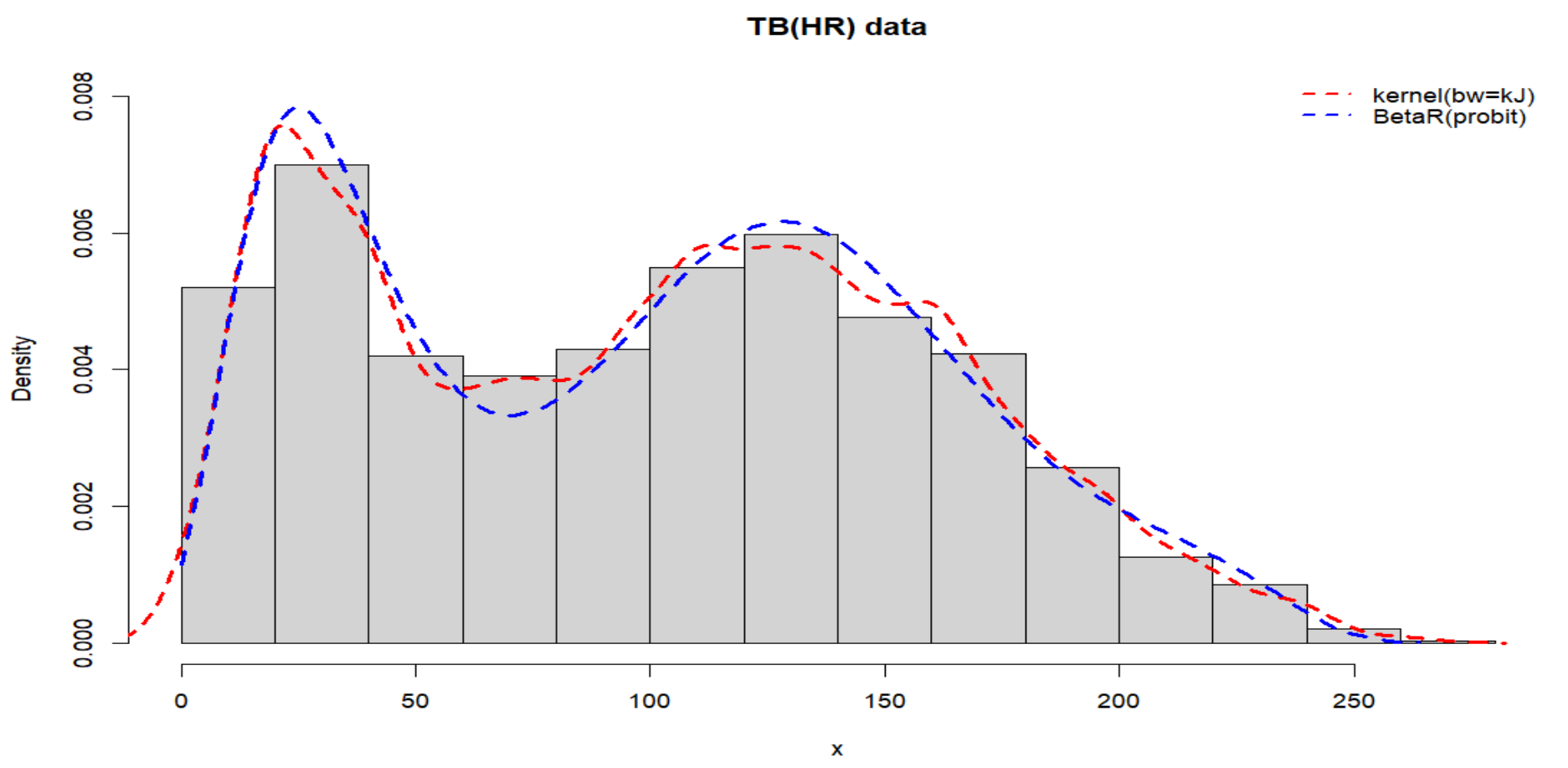

Figure 22. kernel (SJ) and BetaR (probit) estimated density function for 2276 records on baseball team batting

Additional insights into the data can be generated by determining the mode of the given distributions through the numerical derivative method, as implemented in the R-software (2024) package "sfsmisc"; see, Maechler (2024). The modes using BetaR (probit) are 24.5, 70.5 and 128.5.

## 6 Discussion and novelty of the proposed method

The kernel distribution estimation was introduced by Nadaraya (1964a) and is defined by

$$\hat{F}_K(x) = \left(\frac{1}{n}\right)\sum_{i=1}^{n} K\left(\frac{x - X_i}{h}\right)$$

Where $K$ is a given distribution function such as Gaussain or Epanechnikov and $h = h(n) > 0$ $(h \to 0 \; as \; n \to \infty)$ is the bandwidth. While our approach is defined as

$$\hat{F}_j\left(\sum_{i=1}^{n}\sum_{g=1}^{m} b_{g-1}x_i^g\right)$$

$F_j$ could be Gaussian, logistic, extrem value and Cauchy distributions. In kernel smoothing, the parameter $h$ is responsible for achieving smoothness. Conversely, in our estimator, it is the polynomial function with degree $m$ that ensures smoothness. As a result, while the kernel method requires selecting from a broad value of $h > 0$ to attain the desired smoothness, our method necessitates choosing from a much narrower range of $m$ values. Practical experience indicates that a polynomial degree of 3 up to 7 often proves to be appropriate for numerous cases.

With respect to density, the kernel estimators is

$$\hat{f}_K(x) = \left(\frac{1}{n}\right)\sum_{i=1}^{n}\left(\frac{1}{h}\right)K\left(\frac{x-x_i}{h}\right)$$

Our approach is

$$\hat{f}_J(x) = \sum_{i=1}^{n}\sum_{g=1}^{m} g b_{g-1} x_i^{g-1} f_J\left(\sum_{i=1}^{n}\sum_{g=1}^{m} b_{g-1} x_i^{g}\right)$$

The degree of smoothness for a kernel is contingent on the parameter $h$ both inside and outside $K$, necessitating the selection from an extensive array of positive $h$ values to ensure smoothness. Conversely, our method relies on the polynomial and its derivative, both inside and outside the density function. As a result, while the kernel method requires a broad selection of $h$ values to achieve smoothness, our method uses double smoothing polynomial functions that operate within a much narrower range of $m$ and its derivative can be obtained in smooth way. Practical applications suggest that a polynomial degree of 4 or 5 is often adequate. Additionally, our method tends to produce an estimator that is almost unbiased, unlike the kernel method.

Conflict of Interest: The authors do not have any conflicts of interest to declare.